\documentclass[12pt]{JHEP3}
\pdfoutput=1
\usepackage{amsfonts, amsmath, amsthm}
\usepackage{graphicx}

\def \Tr {\mathop{\rm Tr}\nolimits}

\newcommand\ignore[1]{}
\def\one{{\,\hbox{1\kern-.8mm l}}}

\def\a{\alpha}\def\b{\beta}

\def\dslash{\partial\!\!\!/}\def\Dslash{D\!\!\!\!/\,\,}

\newcommand{\Cset}{{\,\,{{{^{_{\pmb{\mid}}}}\kern-.45em{\mathrm C}}}}}

\def \d {\partial}
\newcommand{\be}{\begin{equation}}
\newcommand{\ba}{\begin{eqnarray}}
\newcommand{\bea}{\begin{eqnarray}}

\newcommand{\ee}{\end{equation}}
\newcommand{\eea}{\end{eqnarray}}
\newcommand{\ea}{\end{eqnarray}}

\title{Abelian-Higgs and Vortices from ABJM: towards a string realization of AdS/CMT}
\author{Asadig Mohammed$^{1}$\footnote{asadig@gmail.com}, Jeff Murugan$^{1}$\footnote{jeff@nassp.uct.ac.za} and Horatiu Nastase$^{2}$
\footnote{nastase@ift.unesp.br} \\

$^{1}$The Laboratory for Quantum Gravity \& Strings, \\
Department of Mathematics and Applied Mathematics, \\
University of Cape Town, \\
Private Bag, Rondebosch, 7700, \\
South Africa.\\

$^{2}$Instituto de F\'{i}sica Te\'{o}rica,\\ UNESP-Universidade Estadual Paulista,\\
R. Dr. Bento T. Ferraz 271, Bl. II,\\ S\~ao Paulo 01140-070, SP, Brazil}

\abstract{We present ans\"{a}tze that reduce the mass-deformed ABJM model to gauged Abelian scalar theories, using the fuzzy sphere matrices $G^\alpha$. One such reduction gives a Toda system, for which we find a new type of nonabelian vortex. Another gives the standard Abelian-Higgs model, thereby allowing us to 
embed all the usual (multi-)vortex solutions of the latter into the ABJM model. By turning off the mass deformation at the level of the reduced model, we can also continuously deform to the massive $\phi^4$ theory in the 
massless ABJM case. In this way we can embed the Landau-Ginzburg model into the AdS/CFT correspondence
as a consistent truncation of ABJM. In this context, the mass deformation parameter $\mu$ and a field VEV $\langle\phi\rangle$ act as $g$ and $g_c$ respectively, leading to a well-motivated AdS/CMT
construction from string theory. 
To further this particular point, we propose a simple model for the condensed matter field theory that leads to an approximate description for the ABJM abelianization. Finally, we also find some BPS solutions to the mass-deformed ABJM model with a spacetime interpretation as an M2-brane ending on a spherical M5-brane.
} 

\keywords{ABJM model, abelianization, vortices, M-branes} \preprint{}

\begin{document}

\maketitle

\section{Introduction}

Since its beginnings in 1997, the AdS/CFT correspondence \cite{Maldacena:1997re} 
has found application in a variety of phenomena; not only in quantum gravity but also, increasingly in fields as diverse as low energy QCD and condensed matter. Its original formulation described four dimensional ${\cal N}=4$ Supersymmetric Yang-Mills theory with an $SU(N)$ gauge
group in the large $N$ limit from the perspective of a dual gravitational theory on $AdS_5\times S^5$. As a toy model for the exploration of four dimensional QCD at strong coupling, $\mathcal{N}=4$ SYM, with its large supersymmetry, conformal invariance, and large number of colors $N$ (that ensures that the dual is just a gravitational
theory and not a full string theory) is nearly ideal. By modifying this simple set-up, in particular by breaking 
supersymmetry and conformal invariance, a lot was learned about QCD itself as, for instance, in the Sakai-Sugimoto model \cite{Sakai:2004cn}. A crucial part of this development is that the physics of gauge theories at finite temperature shows remarkable universality, which has translated into applications of ${\cal N}=4$ SYM to the high temperature plasmas at RHIC and the ALICE experiment at the LHC (see for example \cite{Gubser:2009fc} for an extensive review and references). 

With the discovery of the pp-wave/BMN correspondence in 2002 \cite{Berenstein:2002jq} came the realization of the importance of operators with large R-charge to a full string theory (and not just  supergravity) description of the dual. This, in turn, led to the description of spin chains from string theory \cite{Minahan:2002ve} and, more generally, to an understanding of the integrable structures on both sides of the correspondence. In a sense, this was the precursor to the application of the gauge/gravity duality to condensed matter physics. More recently, another important example of AdS/CFT appeared, the so-called ABJM model; a three dimensional ${\cal N}=6$ supersymmetric Chern-Simons gauge theory with the gauge group $SU(N)\times SU(N)$, dual to string theory on $AdS_4\times {\mathbb CP}^3$. This model can be considered as a prototype for strongly coupled theories in three dimesions, in particular for planar condensed matter systems. For instance, in \cite{Myers:2010pk,Huijse:2011hp} it was used to study the relativistically invariant quantum critical phase and compressible Fermi surfaces, respectively. These applications of the $AdS/CFT$ correspondence to condensed matter hinge on the idea that, if physics in AdS is always holographic, then we can consider simple theories in AdS, which {\it should} be dual to some strongly coupled conformal field theories (see e.g. \cite{Hartnoll:2009sz,Herzog:2009xv} for a review). In an overwhelming majority of cases considered, the argument for applying the AdS/CFT duality (and trusting the answers it provided) was {\it universality}. In other words, the variety of theories usually considered contain a small subset of abelian operators dual to a small number of fields in AdS, usually a gauge field, some scalars and perhaps some fermions. On the other hand, the relevant condensed matter models one usually wants to describe is usually abelian to begin with. It is not entirely clear then why we can either: i) focus on a small subset of abelian operators of a large 
$N$ system; or ii) consider an abelian analog of the large $N$ system, which would not have a gravity dual.

A better motivated scenario for such an ``AdS/CMT " correspondence would be if, in a large $N$ field theory with a 
gravity dual, we could identify a {\it consistent truncation} of the (in general, nonabelian) field theory to an abelian 
subset corresponding to the {\it collective dynamics} of a large number of fields, and the 
resulting abelian theory would be a relevant condensed matter 
model. It is toward this end that we explore possible abelian reductions of the ABJM model in this article. 
Our strategy will be to look to the matrices $G^\a$ that characterize the 
``fuzzy funnel" BPS state of pure ABJM and the ``fuzzy sphere" ground state of the 
massive deformation of ABJM (mABJM) since they correspond to a collective motion of 
${\cal O}(N)$ out of ${\cal O}(N^2)$ degrees of freedom. They will play a central role in our abelianization ansatz. We will then argue that this ansatz furnishes a {\it consistent truncation} of mABJM and can be used to identify further (phenomenologically) interesting abelianizations. We then show how these find application in condensed matter physics and, finally, we will explore some BPS solutions 
suggested by the abelian ans\"{a}tze together with their spacetime interpretation. The main ideas about the abelianization and application to AdS/CMT were outlined in the letter \cite{Mohammed:2012gi}, and here we present the full details.

At this point, it is worth elaborating on the idea of consistent truncations and the context in which we will use them in this article.To this end, let's recall previous instances where they have been used commonly. Perhaps the most frequently encountered use of consistent truncations is in the context of AdS/CFT. Here, one usually deals with 
a {\it classical theory} on the gravity side, so the existence of a consistent truncation to some reduced theory means that we can safely drop the other modes, since they will appear only in quantum loops. However, another common utilization of truncations is in supergravity compactifications. Here the relevant question is whether or not we can safely retain only the 
phenomenologically interesting reduced four-dimensional theory. 
If there exists a consistent truncation, one can check whether couplings to the ``nonzero modes" 
can be made arbitrarily small or whether, more commonly, the masses of the nonzero modes are much larger than the mass parameters of the reduced theory. We will argue that it is the latter usage of consistent truncations that is relevant in our case and as a result, at low energies we can, with no loss of physics, drop the nonzero modes even from the quantum theory. 

Another issue that warrants clarification is our use of the {\it collective dynamics} of ${\cal O}(N)$ modes. To see why this is different from the dynamics of any {\it single} field, consider a large number, $N$, of branes in some gravitational background. 
A classical solution obtained by turning on fields in all $N$ branes corresponds, in the gravity dual, to a finite deformation of the background, 
in stark contrast to turning on fields on a {\it single brane} which does not produce a finite effect in the dual background.

The rest of this paper is organized as follows. In section 2 we explore general abelianization 
ans\"{a}tze involving $G^\a$, and identify two important 
cases of further consistent truncations for this model. In section 3 we study one 
of them, which, for BPS states, leads to a Toda system that possesses 
vortex-type solutions with topological charge and finite energy, but with 
$|\phi|\rightarrow 0$ at both $r\rightarrow 0$ and $r\rightarrow \infty$, 
which we describe numerically. In section 4 we describe a second case, 
more relevant for the AdS/CMT motivation above and find a reduction that, 
depending on certain parameters, gives us either an abelian-Higgs model, or a $\phi^4$ 
(relativistic Landau-Ginzburg) theory. 
In section 5 we study the relevance of this reduction for condensed matter and 
AdS/CMT and sketch a simple condensed matter model that reproduces 
the general features of abelianization. In section 6 we study some BPS solutions 
suggested by the abelianizations. Finally, in section 7, we provide 
a possible spacetime interpretation for these solutions in terms of M2-branes on a background spacetime.
\section{ABJM, massive ABJM and their Truncations}

The ABJM model \cite{Aharony:2008ug} is obtained as the IR limit of the theory of $N$ coincident M2-branes moving in $\mathbb R^{2,1}\times \mathbb C^4/\mathbb Z_k$. 
It is a ${\cal N}=6$ supersymmetric $U(N)\times U(N)$ Chern-Simons gauge theory at level $(k,-k)$, with bifundamental scalars $C^I$ and fermions $
\psi_I$, $I=1,...,4$ in the fundamental of the $SU(4)_R$ symmetry group. The gauge fields are denoted by $A_\mu$ and $\hat A_\mu$. 
Its action is given by
\bea
S&=&\int d^3x \Bigg( \frac{k}{4\pi}\epsilon^{\mu\nu\lambda}\Tr\left(A_{\mu}\partial_{\nu}A_{\lambda}+\frac{2i}{3}A_{\mu}A_{\nu}A_{\lambda}-\hat{A}_{\mu}\partial_{\nu}\hat{A}_{\lambda}
      -\frac{2i}{3}\hat{A}_{\mu}\hat{A}_{\nu}\hat{A}_{\lambda}\right)\cr
      &-&\Tr D_{\mu}C_{I}^{\dagger}D^{\mu}C^{I}
      -i\Tr \psi^{I\dagger}\gamma^{\mu}D_{\mu}\psi_{I}+\frac{4\pi^2}{3k^2}\Tr \Big(C^{I}C_{I}^{\dagger}C^{J}C_{J}^{\dagger}C^{K}C_{K}^{\dagger}\cr
&+&C_{I}^{\dagger}C^{I}C_{J}^{\dagger}C^{J}C_{K}^{\dagger}C^{K}
      +4C^{I}C_{J}^{\dagger}C^{K}C_{I}^{\dagger}C^{J}C_{K}^{\dagger}
      -6C^{I}C_{J}^{\dagger}C^{J}C_{I}^{\dagger}
      C^{K}C_{K}^{\dagger}C^{K}\Big)\cr
&+&\frac{2\pi i}{k}\Tr\Big(C_{I}^{\dagger}C^{I}\psi^{J\dagger}\psi_{J}-\psi^{J\dagger}C^{I}C_{I}^{\dagger}\psi_{J}
      -2C_{I}^{\dagger}C^{J}\psi^{I\dagger}\psi_{J}
      +2\psi^{J\dagger}C^{I}C_{J}^{\dagger}\psi_{J}\cr
&+&\epsilon^{IJKL} C_{I}^{\dagger}\psi_{J}C_{K}^{\dagger}\psi_{L}-\epsilon_{IJKL}C^{I}\psi^{J\dagger}C^{K}\psi^{L\dagger}\Big)
      \Bigg)
\eea
where the gauge-covariant derivative is 
\be
      D_{\mu} C^{I}=\partial_{\mu}C^{I}+iA_{\mu}C^{I}-iC_{I}\hat{A}_{\mu}.
\ee
The action has a $SU(4)\times U(1)$ R-symmetry associated with the ${\cal N}=6$ supersymmetries. 

There is a maximally supersymmetric ({\em i.e.}, preserving all ${\cal N}=6$) massive deformation of the model with a parameter $\mu$ \cite{Gomis:2008vc,Terashima:2008sy}, 
which breaks the R-symmetry down to $SU(2) \times SU(2)\times U(1)_{A}\times U(1)_{B}\times \mathbb{Z}_{2}$ by splitting the scalars as
\be
     C^{I}=(Q^{\alpha},R^{\alpha}); \qquad \alpha=1,2
\ee
The $\mathbb{Z}_{2}$ action swaps the matter fields $Q^{\alpha}$ and $R^{\alpha}$, while the $SU(2)$ factors act individually on the doublets ${Q^{\alpha}}$ and ${R^{\alpha}}$ respectively and $U(1)_{A}$ symmetry rotates $Q^{\alpha}$ with a phase $+1$  and $R^{\alpha}$ with a phase $-1$. The mass deformation, besides giving a mass to the fermions, changes the potential of the theory. The bosonic part of the deformed action can 
be written as 
\bea
  \mathcal{L}_{\rm{Bosonic}}&=&\frac{k}{4\pi}\epsilon^{\mu \nu \lambda}\Tr\left(A_{\mu}\partial_{\nu}   
  A_{\lambda}+\frac{2i}{3}A_{\mu}A_{\nu}A_{\lambda}-\hat{A}_{\mu}\partial_{\nu}\hat{A}_{\lambda}
  -\frac{2i}{3}\hat{A}_{\mu}\hat{A}_{\nu}\hat{A}_{\lambda}\right)\nonumber\\
  &-&\Tr|D^{\mu}Q^{\alpha}|^2-\Tr|D^{\mu}R^{\alpha}|^2-V
  \label{mass deformed ABJM lagrangian}
\eea
where the potential is
\be
     V=\Tr\left(|M^{\alpha}|^2+|N^{\alpha}|^2\right),
\ee
and where
\bea
  M^{\alpha}&=& \mu Q^{\alpha}+\frac{2\pi}{k}\Big(2Q^{[\alpha}Q^{\dagger}_{\beta}Q^{\beta]}+R^{\beta}  
  R^{\dagger}_{\beta}Q^{\alpha}-Q^{\alpha}R^{\dagger}_{\beta}R^{\beta}
  +2Q^{\beta}R^{\dagger}_{\beta}R^{\alpha}-2R^{\alpha}R^{\dagger}_{\beta}Q^{\beta}\Big),\nonumber\\
  N^{\alpha} &=& -\mu R^{\alpha}+\frac{2\pi}{k}\Big(2R^{[\alpha}R^{\dagger}_{\beta}R^{\beta]}+Q^{\beta}  
  Q^{\dagger}_{\beta}R^{\alpha}-R^{\alpha}Q^{\dagger}_{\beta}Q^{\beta}
  +2R^{\beta}Q^{\dagger}_{\beta}Q^{\alpha}-2Q^{\alpha}Q^{\dagger}_{\beta}R^{\beta}\Big).\cr
  &&\label{mandn}
\eea
The equations of motion of the bosonic Lagrangian (\ref{mass deformed ABJM lagrangian}) are
\bea
  D_{\mu}D^{\mu}Q^{\alpha}&=&\frac{\partial V}{\partial Q^{\dagger}_{\alpha}},\hspace{1cm}
  D_{\mu}D^{\mu}R^{\alpha}=\frac{\partial V}{\partial R^{\dagger}_{\alpha}},\nonumber\\
  F_{\mu\nu}&=&\frac{2\pi }{k}\epsilon_{\mu\nu\lambda}J^{\lambda},\hspace{1cm}
  \hat{F}_{\mu\nu}=\frac{2\pi }{k}\epsilon_{\mu\nu\lambda}\hat{J}^{\lambda},
  \label{EulerLagrangeeqns}
\eea
where the field strength $F_{\mu \nu}=\partial_{\mu}A_{\nu}-\partial_{\nu}A_{\mu}+i[A_{\mu},A_{\nu}]$ and the two gauge currents $J^{\mu}$ and $\hat{J}^{\mu}$, given by
\bea
  J^{\mu}&=&i\left(Q^{\alpha}(D^{\mu}Q^{\alpha})^{\dagger}-(D^{\mu}Q^{\alpha})Q_{\alpha}^{\dagger}  
  +R^{\alpha}(D^{\mu}R^{\alpha})^{\dagger}-(D^{\mu}R^{\alpha})R_{\alpha}^{\dagger}\right),\nonumber\\
  {}\\
  \vspace{-2cm}
  \hat{J}^{\mu}&=&-i\left(Q_{\alpha}^{\dagger}(D^{\mu}Q^{\alpha}-(D^{\mu}Q^{\alpha})^{\dagger}Q^{\alpha}
  +R_{\alpha}^{\dagger}(D^{\mu}R^{\alpha}-(D^{\mu}R^{\alpha})^{\dagger}R^{\alpha} \right),\nonumber
  \label{abelian current}
\eea
are covariantly conserved so that $\nabla_{\mu}J^{\mu}=\nabla_{\mu}\hat J^\mu=0$. In addition, there are two abelian currents $j^{\mu}$  and $\hat{j}^{\mu}$ corresponding to the global $U(1)_{A}$ and $U(1)_{B}$ invariances, given by
\bea
   j^{\mu}&=&i\Tr\left(Q^{\alpha}(D^{\mu}Q^{\alpha})^{\dagger}-(D^{\mu}Q^{\alpha})Q_{\alpha}^{\dagger}  
  +R^{\alpha}(D^{\mu}R^{\alpha})^{\dagger}-(D^{\mu}R^{\alpha})R_{\alpha}^{\dagger}\right),\nonumber\\ 
  {}\\
  \hat{j}^{\mu}&=-&i\Tr\left(Q_{\alpha}^{\dagger}(D^{\mu}Q^{\alpha})-(D^{\mu}Q^{\alpha})^{\dagger}Q^{\alpha}  
  +R_{\alpha}^{\dagger}(D^{\mu}R^{\alpha})-(D^{\mu}R^{\alpha})^{\dagger}R^{\alpha}\right),\nonumber
\eea
which are ordinarily conserved {\it i.e.} $\partial_{\mu}j^{\mu}=\d_\mu \hat j^\mu=0$. By choosing the gauge $A_0=\hat A_0=0$, the energy density (Hamiltonian) is given by\footnote{Note that the terms $A_1\dot A_2-A_2\dot A_1$ cancel from $p\dot q-L$, and the rest of the CS term involve $A_0$} 
\bea 
  H&=&\Tr\left[(D_0Q^{\alpha})^{\dagger}(D_0Q^{\alpha})+(D_iQ^{\alpha})^{\dagger}(D_iQ^{\alpha})\right.\cr
  &&\left.+(D_0R^{\alpha})^{\dagger}(D_0R^{\alpha})
  +(D_iR^{\alpha})^{\dagger}(D_iR^{\alpha})+V\right].
  \label{Hamiltonian}
\eea
Since this is a Chern-Simons theory, the equations of motion must be supplemented with the 
Gauss law constraints
\bea
  F_{12}&=&\frac{2\pi i}{k}J^{0}=\frac{2\pi i}{k}\left(Q^{\alpha}(D^{0}Q^{\alpha})^{\dagger}-(D^{0}  
  Q^{\alpha})Q_{\alpha}^{\dagger}+R^{\alpha}(D^{0}R^{\alpha})^{\dagger}-(D^{0}R^{\alpha})R_{\alpha}^{\dagger}  
  \right),\nonumber\\
  {}\\
  \hat{F}_{12}&=&\frac{2\pi i}{k}\hat{J}^{0}=-\frac{2\pi i}{k}\left(Q_{\alpha}^{\dagger}(D^{0}Q^{\alpha})-(D^{0}  
  Q^{\alpha})^{\dagger}Q^{\alpha}+R_{\alpha}^{\dagger}(D^{0}R^{\alpha})-(D^{0}R^{\alpha})^{\dagger}R^{\alpha}  
  \right).\nonumber
  \label{gaussconstraint}
\eea
The gauge choice is not as restrictive as it would seem. Choosing (as we do below for our abelianization) $A_0$ and $\hat A_0$ different from zero produces an extra term in the Hamiltonian of the form $\epsilon^{\mu\nu\lambda}\Tr[A_\mu A_\nu A_\lambda-\hat A_\mu\hat A_\nu \hat A_\lambda]$. In the abelian case this vanishes anyway since it is proportional to $\epsilon^{\mu\nu\lambda}a_\mu^{(i)}a_\nu^{(j)}a_\lambda^{(k)}$ and there
are only two $a^{(i)}_\mu$'s. So in the abelian case, the Hamiltonian is the same even away from the gauge $A_0=\hat A_0=0$. 
The mass deformed theory has ground states of the fuzzy sphere type given by 
\be
R^\a=c G^\a;\;\;\; Q^\a=0\;\;\;{\rm and}\;\;\;
Q^\dagger_\a=c G^\a;\;\;\; R^\a=0
\ee
where $c\equiv\sqrt{\frac{\mu k}{2\pi}}$ and the matrices $G^\a$, $\a=1,2$, satisfy the equations 
\cite{Gomis:2008vc,Terashima:2008sy}
\be
G^\a=G^\a G^\dagger_\b G^\b-G^\b G^\dagger_\b G^\a.
\ee
It was shown in \cite{Nastase:2009ny,Nastase:2010uy} that this solution corresponds to a fuzzy 2-sphere. 

An explicit solution of these equations is given by 
\bea\label{BPSmatrices}
&& ( G^1)_{m,n }    = \sqrt { m- 1 } ~\delta_{m,n}\,, \cr
&& ( G^2)_{m,n} = \sqrt { ( N-m ) } ~\delta_{ m+1 , n }\,, \cr
&& (G_1^{\dagger} )_{m,n} = \sqrt { m-1} ~\delta_{m,n}\,, \cr
&& ( G_2^{\dagger} )_{m,n} = \sqrt { (N-n ) } ~\delta_{ n+1 , m }\,.
\eea
Clearly, these matrices satisfy $G^1=G^\dagger_1$ also. 
In the case of the pure ABJM, there is a {\em BPS solution} of the fuzzy funnel type with $c$ replaced by
\be
c(s)=\sqrt{\frac{k}{4\pi s}}\,,
\ee
instead, where $s$ is one of the two spatial coordinates of the ABJM model. The matrices $G^\a$ are bifundamental under $U(N)\times U(N)$, therefore $G^1G^\dagger_1$ and $G^2 G^\dagger_2$ are in the adjoint of the first 
$U(N)$, and $G^\dagger_1 G^1$ and $G^\dagger_2 G^2$ are in the adjoint of the second. 

\subsection{An Abelianization Ansatz}
Given all these properties of the $G^{\alpha}$ matrices, it is reasonable to choose the following abelianization ansatz
\bea
A_{\mu}&=&a^{(2)}_{\mu}G^{1}G_{1}^{\dagger}+a^{(1)}_{\mu}G^{2}G_{2}^{\dagger}\,,\nonumber\\
\hat{A}_{\mu}&=&a^{(2)}_{\mu}G_{1}^{\dagger}G^{1}+a^{(1)}_{\mu}G_{2}^{\dagger}G^{2}\,,\nonumber\\
Q^{\alpha}&=&\phi_{\alpha}G^{\alpha}\nonumber\,,\\
R^{\alpha}&=&\chi_{\alpha}G^{\alpha}\,,
\eea
with no summation over $\a$ in the ansatz for $Q^\a, R^\a$;
$a_{\mu}^{(1)}$ and $a_{\mu}^{(2)}$ real-valued vector fields and $\phi_{\alpha}$, $\chi_{\alpha}$ complex-valued scalar fields. 

Since $G^{1}G_{1}^{\dagger}$ commutes with $G^{2}G_{2}^{\dagger}$ and $G_{1}^{\dagger}G^{1}$ commutes with $G_{2}^{\dagger}G^{2}$,
the gauge fields $a_{\mu}^{(i)}$ are abelian and the field strengths decompose as
\bea
  F_{\mu\nu}=\partial_{\mu}A_{\nu}-\partial_{\nu}A_{\mu}+i[A_{\mu},A_{\nu}]=
  f_{\mu\nu}^{(2)}G^{1}G_{1}^{\dagger}+f_{\mu\nu}^{(1)}G^{2}G_{2}^{\dagger}\,,\nonumber\\
  \\
  \hat{F}_{\mu\nu}=\partial_{\mu}\hat{A}_{\nu}-\partial_{\nu}\hat{A}_{\mu}+i[\hat{A}_{\mu},
  \hat{A}_{\nu}] =f_{\mu\nu}^{(2)}G_{1}^{\dagger}G^{1}+f_{\mu\nu}^{(1)}G_{2}^{\dagger}G^{2}\,,\nonumber
\eea
with the {\it abelian} field strengths $ f_{\mu\nu}^{(i)}=\partial_{\mu}a^{(i)}_{\nu}-\partial_{\nu}a^{(i)}_{\mu}$.

With this ansatz, the Chern-Simons term becomes
\be
  -\frac{k}{4\pi}\frac{N(N-1)}{4}\epsilon^{\mu \nu \lambda}\Big(a^{(2)}_{\mu}f^{(1)}_{\nu \lambda}
  +a^{(1)}_{\mu}f^{(2)}_{\nu \lambda}\Big),
\ee
while the covariant derivatives $D_\mu Q^\a$ and $ D_\mu R^\a$ give rise to 
\bea
  D_{\mu}\phi_{i}&=&(\partial_{\mu}-ia_{\mu}^{(i)})\phi_{i}\,,\nonumber\\
  \\
  D_{\mu}\chi_{i}&=&(\partial_{\mu}-ia_{\mu}^{(i)})\chi_{i}\,,\nonumber
\eea
and the values for $M^\a, N^\a$ are given by
\bea
  M^{1}&=&\frac{2\pi}{k}\left[\phi_{1}\big(c^{2}+|\phi_{2}|^{2}-|\chi_{2}|^{2}\big)-2\chi_{1}\overline{\chi}  
  _{2}\phi_{2}\right]G^{1}\,,\nonumber\\
  M^{2}&=&\frac{2\pi}{k}\left[\phi_{2}\big(c^{2}+|\phi_{1}|^{2}-|\chi_{1}|^{2}\big)-2\overline{\chi_{1}}  
  \chi_{2}\phi_{1}\right]G^{2}\,,\nonumber\\
  \\
  N^{1}&=&\frac{2\pi}{k}\left[\chi_{1}\big(-c^{2}+|\chi_{2}|^{2}-|\phi_{2}|^{2}\big)-2\phi_{1}\overline{\phi_{2}}  
  \chi_{2}\right]G^{1}\,,\nonumber\\
  N^{2}&=&\frac{2\pi}{k}\left[\chi_{2}\big(-c^{2}+|\chi_{1}|^{2}-|\phi_{1}|^{2}\big)-2\overline{\phi}  
  _{1}\phi_{2}\chi_{1}\right]G^{2}\nonumber
\eea
where as before, $c^2=\mu k/(2\pi)$. Substituting into the potential gives 
\bea
  V&=&\frac{2\pi^{2}}{k^2}N(N-1)\Big[(|\phi_{1}|^{2}+|\chi_{1}|^{2})\big(|\chi_{2}|^{2}-|\phi_{2}|^{2}
  -c^{2}\big)^{2}\nonumber\\
  &+&(|\phi_{2}|^{2}+|\chi_{2}|^{2})\big(|\chi_{1}|^{2}-|\phi_{1}|^{2}-c^{2}\big)^{2}\label{abelianpot}\\
  &+&4|\phi_{1}|^{2}|\phi_{2}|^{2}(|\chi_{1}|^{2}+|\chi_{2}|^{2})+4|\chi_{1}|^{2}|\chi_{2}|^{2}(|\phi_{1}|^{2}+|  
  \phi_{2}|^{2})\Big]\,.\nonumber
\eea
Note that the interchange of $\chi$ with $\phi$ (which changes $Q^\a$ with $R^\a$) is equivalent to a 
change in the sign of $c^2$, {\it i.e.} either a change in the sign of $\mu$, or of $k$. Putting everything together then gives the final {\it abelian} effective action
\be 
  S=-\frac{N(N-1)}{2}\int d^{3}x\Bigg[\frac{k}{4\pi}\epsilon^{\mu \nu \lambda}\big(a^{(2)}_{\mu}f^{(1)}_{\nu   
  \lambda}+a^{(1)}_{\mu}f^{(2)}_{\nu \lambda}\big)
  +|D_{\mu}\phi_{i}|^{2}+|D_{\mu}\chi_{i}|^{2}+U(|\phi_{i}|,|\chi_{i}|)\Bigg]\,,
  \label{abelianmaster}
\ee
with a rescaled potential $U\equiv 2V/N(N-1)$. Since the effective theory derives from a Chern-Simons theory, the equations of motion need to be supplemented with the Gauss law constraints which, in our ansatz, reduce to 
\bea
  f_{12}^{(2)}&=&\frac{2\pi i}{k}[\phi_1(D^0\phi_1)^\dagger-(D^0\phi_1)\phi_1^\dagger+\chi_1(D^0\chi_1)^\dagger-  
  (D^0\chi_1)\chi_1^\dagger]\,,\nonumber\\
  \\
  f_{12}^{(1)}&=&\frac{2\pi i}{k}[\phi_2(D^0\phi_2)^\dagger-(D^0\phi_2)\phi_2^\dagger+\chi_2(D^0\chi_2)^\dagger-  
  (D^0\chi_2)\chi_2^\dagger]\,,\nonumber
\eea
We see, however, that these are nothing but the $a_0^{(1)},a_0^{(2)}$ equations of motion for the action (\ref{abelianmaster}). As we will need to work away from the $a_0^{(1)}=a_0^{(2)}=0$ gauge, we don't need to impose them. 

\subsection{Consistent Truncations}
A key point to note about this abelianization ansatz is that it a {\it consistent truncation} of the original ABJM theory in the sense that, using the facts that $M^\a\propto G^a$, $N^\a\propto G^\a$,
$D_\mu D^\mu(\phi_\a G^\a)=(D_\mu D^\mu \phi_a)G^\a$ and
$D_\mu D^\mu(\chi_\a G^\a)=(D_\mu D^\mu \chi_a)G^\a$, the equations of motion that follow from the 
action (\ref{abelianmaster}), 
\bea
  \frac{k}{4\pi}\epsilon^{\mu\nu\lambda}f^{(1)}_{\mu\nu}&=&i\Big[\overline{\phi}_{2}D^{\lambda}\phi_{2}- 
  \phi_{2}\overline{D^{\lambda}\phi_{2}}
  +\overline{\chi}_{2}D^{\lambda}\chi_{2}-\chi_{2}\overline{D^{\lambda}\chi_{2}}\Big]\,,\nonumber\\
  \\
  \frac{k}{4\pi}\epsilon^{\mu\nu\lambda}f^{(2)}_{\mu\nu}&=&i\Big[\overline{\phi}_{1}D^{\lambda}\phi_{1}- 
  \phi_{1}\overline{D^{\lambda}\phi_{1}}
  +\overline{\chi}_{1}D^{\lambda}\chi_{1}-\chi_{1}\overline{D^{\lambda}\chi_{1}}\Big]\,,\nonumber 
  \label{equations of motion of the gauge fields} 
\end{eqnarray}  
and
\begin{eqnarray}
  &&D_{\mu}D^{\mu}\phi_{1}\cr
  &=&\frac{4\pi^{2}}{k^2}\Big[\big(|\chi_{2}|^{2}-|\phi_{2}|^{2}-c^{2}\big)^{2}+2\big(|  
  \phi_{2}|^{2}+|\chi_{2}|^{2}\big)\big(|\phi_{1}|^{2}
  +|\chi_{1}|^{2}+c^{2}\big)+4|\phi_{2}|^{2}|\chi_{2}|^{2}\Big]\phi_{1}\,,\nonumber\\     
  &&D_{\mu}D^{\mu}\phi_{2}\cr
  &=&\frac{4\pi^{2}}{k^2}\Big[\big(|\chi_{1}|^{2}-|\phi_{1}|^{2}-c^{2}\big)^{2}+2\big(|  
  \phi_{1}|^{2}+|\chi_{1}|^{2}\big)
  \big(|\phi_{2}|^{2}+|\chi_{2}|^{2}+c^{2}\big)+4|\phi_{1}|^{2}|\chi_{1}|^{2}\Big]\phi_{2}\,,\nonumber\\ 
  &&D_{\mu}D^{\mu}\chi_{1}\cr
  &=&\frac{4\pi^{2}}{k^2}\Big[\big(|\chi_{2}|^{2}-|\phi_{2}|^{2}-c^{2}\big)^{2}+2\big(|  
  \phi_{2}|^{2}+|\chi_{2}|^{2}\big)\big(|\phi_{1}|^{2}+|\chi_{1}|^{2}-c^{2}\big)+4|\phi_{2}|^{2}|\chi_{2}|  
  ^{2}\Big]\chi_{1}\,,\nonumber\\ 
  &&D_{\mu}D^{\mu}\chi_{2}\cr
  &=& \frac{4\pi^{2}}{k^2}\Big[\big(|\chi_{1}|^{2}-|\phi_{1}|^{2}-c^{2}\big)^{2}+2\big(|  
  \phi_{1}|^{2}+|\chi_{1}|^{2}\big)\big(|\phi_{2}|^{2}+|\chi_{2}|^{2}-c^{2}\big)+4|\phi_{1}|^{2}|\chi_{1}|^{2}\Big]  
  \chi_{2}\,,\cr
  &&
\eea
satisfy the higher original ABJM equations of motion (\ref{EulerLagrangeeqns}) and Gauss constraints (\ref{gaussconstraint}).

Since $\Tr[G^1G_1^\dagger]=\Tr[G_1^\dagger G_1]=\Tr[G^2G^\dagger_2]=\Tr[G^\dagger_2 G^2]=N(N-1)/2$, the energy density (Hamiltonian) is 
\be
H=\frac{N(N-1)}{2}[|D_0\phi_i|^2+|D_0\chi_i|^2+|D_a\phi_i|^2+|D_a\chi_i|^2]+V
\ee
where $a,b=1,2$. Note also that away from the gauge $A_0=\hat A_0=0$ (which imply that $a_0^{(i)}=0$), we would, in principle, have a term cubic in the gauge fields in the 
Hamiltonian. This however vanishes in the abelian case, so the above result is correct in general. This abelianization,  with its four complex scalar fields, is rather general. We will study further reductions of it involving only two scalars. Looking at the scalar equations of motion above we see that putting any two of the scalars to zero is again a consistent truncation. 

\begin{itemize}
  \item
  A trivial choice turns out to be $\chi_2=\phi_2=0$ (or equivalently $\chi_1=\phi_1=0$), since in that case, the   
  potential reduces to a simple mass term, 
  \be
    V=\frac{4\pi^2c^4}{k^2}(|\phi_1|^2+|\chi_1|^2)
  \ee
  while at the same time, the only $a^{(2)}_\mu$ dependence remains in the Chern-Simons term, 
  $\sim \int \epsilon a^{(2)}  
  f^{(1)}$, so its equation of motion is $f^{(1)}_{\mu\nu}=0$, which means $a_\mu^{(1)}$ is also trivial (pure   
  gauge). So we remain with two massive complex scalar fields coupled to one trivial gauge field (pure gauge, with no   
  kinetic term), an uninteresting model.
  \item
  A much more interesting choice is $\phi_1=\phi_2=0$, which will turn out to lead (with some modifications) 
  to the Abelian-Higgs model. Since we will study this separately and extensively in section 4, we will not discuss 
  it further here. 
  \item
  Finally, setting $\chi_1=\phi_2=0$, and renaming $\chi_2$ to $\phi_2$ for simplicity, we get
  \bea
    S&=&-\frac{N(N-1)}{2}\int d^{3}x\Bigg[\frac{k}{4\pi}\epsilon^{\mu \nu \lambda}\Big(a^{(2)}_{\mu}f^{(1)}  
    _{\nu \lambda}+a^{(1)}_{\mu}
    f^{(2)}_{\nu \lambda}\Big)+|D_{\mu}\phi_{i}|^{2}+U(|\phi_{i}|,|\chi_{i}|)\Bigg]\nonumber\\
    \label{newhiggs}\\
    V&=&\frac{2\pi^2}{k^2}N(N-1)[|\phi_1|^2(|\phi_2|^2-c^2)^2+|\phi_2|^2(|\phi_1|^2+c^2)^2]\,,\nonumber
  \eea
  and energy density (Hamiltonian)
  \be
    H=\frac{N(N-1)}{2}[|D_0\phi_i|^2+|D_a\phi_i|^2]+V\,.
  \ee
\end{itemize}
We should note that, until now, we have worked only with the {\it massive} ABJM model, but that we can analyze the {\it massless} (or pure) ABJM model in a straightforward way by setting $c=0$. Since $c$ 
appears only in the potential, we can check that the model with potential (\ref{abelianpot}) is symmetric under interchange of $\phi_i\leftrightarrow
\chi_i$. For the model with $\chi_1=\phi_2=0$ above, for example, we obtain a purely quartic potential,
\be
  V=\frac{2\pi ^2}{k^2}N(N-1)|\phi_1|^2|\phi_2|^2(|\phi_1|^2+|\phi_2|^2)\,.
\ee

\section{New vortex solutions for a Toda system}

We now study BPS solutions of the effective model (\ref{newhiggs}). Before doing so, it is worth taking a step back, and considering the more general case of the $Q^2=R^1=0$
reduction, with only $Q^1=Q$ and $R^2=R$ nonzero, but without the abelianization ansatz. There we can `complete squares' in the Hamiltonian density and 
write it, in complete analogy to the usual Abelian-Higgs model, as 
\bea
\mathcal{H}&=&\Tr|D_{0}Q-iM|^2+\Tr|D_{0}R+iN|^2+\Tr|D_{-}Q|^2+\Tr|D_{+}R|^2\cr
&+&i\epsilon^{ab}\partial_{a}\Tr\Big(Q^{\dagger}(D_{b}Q)-R^{\dagger}(D_{b}R)\Big) +\mu j_{0}\,,
\eea
where $\mu j_0=\mu k/(2\pi) \Tr\left(F_{12}\right)$ and $D_{\pm}\equiv D_1\pm iD_2$. 
Just as in the Abelian-Higgs model, the term on the second line (with $\epsilon^{ab}$) is 
zero on the configurations of interest, since 
\be
  \int_V d^2x\, \epsilon^{ab}\d_a\left(\phi^\dagger D_b \phi\right)=\int_{\d V_\infty}\left(\phi^\dagger 
  D_a\phi\right)\,dx_a
\ee
and $D_a\phi\rightarrow 0$ at $r\rightarrow \infty$ for $\phi=Q,R$ in order to have finite energy configurations.
Moreover, the perfect squares on the first line are minimized by the BPS equations
\be
D_-Q=0;\;\;\;
D_+R=0;\;\;\;
D_0Q=iM;\;\;\;
D_0R=-iN\,,
\ee
which leaves just the {\it topological term}, $\mu j_0$. The BPS equations together with the Gauss law constraints are 
\bea
  D_{-}Q&=&0\,,\nonumber\\
  D_{+}R&=&0\,, \nonumber\\
  D_{0}Q&=&i\mu Q-\frac{2\pi i}{k}\Big[QR^{\dagger}R-RR^{\dagger}Q\Big]\,,\nonumber\\
  \\
  D_{0}R&=&i\mu R+\frac{2\pi i}{k}\Big[RQ^{\dagger}Q-QQ^{\dagger}R\Big]\,,\nonumber\\
  F_{12}&=& -\frac{4\pi\mu}{k}(QQ^{\dagger}+RR^{\dagger})+\frac{8\pi^2}{k^2}\Big[QR^{\dagger}RQ^{\dagger}-  
  RQ^{\dagger}QR^{\dagger}\Big]\,,\nonumber\\     
  \hat{F}_{12}&=&-\frac{4\pi\mu}{k}(Q^{\dagger}Q+R^{\dagger}R)+\frac{8\pi^2}{k^2}\Big[R^{\dagger}  
  QQ^{\dagger}R-Q^{\dagger}RR^{\dagger}Q\Big]\,,\nonumber
  \label{self-dualChernSimonsequations}
\eea
where, in the Gauss law constraints, we have already substituted the BPS equations for $D_0Q,D_0R$.
These equations are more general, and can be used in principle to find nonabelian BPS solutions. In practice, they are still too difficult to solve analytically so from now on we will go back to the abelian case $Q=\phi_1G^1, R=\phi_2
G^2$. There, the BPS equations for $D_0Q$ and $D_0R$ become
\bea
  (\d_0-ia_0^{(1)})\phi_1&=&-\frac{2\pi i}{k}\phi_1\left[|\phi_2|^2-\frac{\mu k}{2\pi}\right]\,,\nonumber\\
  \\
  (\d_0-ia_0^{(2)})\phi_2&=&\frac{2\pi i}{k}\phi_2\left[|\phi_1|^2+\frac{\mu k}{2\pi}\right]\,.\nonumber
\eea
For static configurations (for which $\d_0\phi_i=0$) these equations can be solved for $a_0^{(i)}$ as
\bea
a_0^{(1)}&=&\frac{2\pi}{k}\left[|\phi_2|^2-\frac{\mu k}{2\pi}\right]\,,\nonumber\\
\\
a_0^{(2)}&=&-\frac{2\pi}{k}\left[|\phi_1|^2+\frac{\mu k}{2\pi}\right]\,.\nonumber
\eea
In other words, the $a_{0}^{(i)}$ are completely specified by the scalar fields $\phi_{i}$ and spatial components of the abelian gauge fields $a_{a}^{(i)}$, $a=1,2$.  
Consequently, the (temporal) gauge $a_0^{(i)}=0$ would be inconsistent with the BPS equations. This is different from the Abelian-Higgs model, where 
the one can set both $a_0=0$ and $\d_0=0$, reducing the system to a two dimensional one (for the spatial components). Here, this would be inconsistent
with the Gauss law constraint which, for a Chern-Simons gauge field, relates $F_{12}$ to terms with $D_0Q$ and $D_0R$ so that, if $F_{12}$ is nonzero, so too is $D_0Q$ and $D_0R$.
Finally, the Gauss law constraints in the BPS case reduce to 
\bea
  f_{12}^{(1)}&=&-\frac{8\pi^2}{k^2}|\phi_2|^2\left(|\phi_1|^2+\frac{\mu k}{2\pi}\right)\,,\nonumber\\
  \\
  f_{12}^{(2)}&=&\frac{8\pi^2}{k^2}|\phi_1|^2\left(|\phi_2|^2-\frac{\mu k}{2\pi}\right)\,.\nonumber
  \label{bpsgauss}
\eea
In order to facilitate the rest of the analysis of the BPS system, it will prove useful to complexify the $(x_{1},x_{2})-$plane and write $z=x^1+ix^2$. As usual, this induces a complexification of the derivatives as well as the gauge fields as, 
\be
  \d=\frac{1}{2}(\d_1-i\d_2)\,,\;\;\;
  a^{(i)}\equiv a^{(i)}_z=\frac{1}{2}(a^{(i)}_1-ia^{(i)}_2)\,,\nonumber
\ee
together with their complex conjugates. This, in turn, implies that $f_{12}^{(i)}=-2i[\d \bar a^{(i)}-$ $\bar \d a^{(i)}]$, so that the BPS equations $D_-Q=D_+R=0$ become simply 
\be
  \d\phi_1-ia^{(1)}\phi_1=0,\;\;\;
  \bar\d\phi_2-i\bar a^{(2)}\phi_2=0\,.
  \label{bpsvortex}
\ee
Equations (\ref{bpsvortex}) together with the Gauss law constraints constitute a complete set which, as we argue below, possess at least one simple set of finite energy, spatially localized solutions of the vortex type {\it i.e. isolated zeros of the (complex) scalar fields with 
nonvanishing winding number}. The analysis follows the same general logic as for the Nielsen-Olesen vortex and we start by writing
\be
\phi_1=|\phi_1|e^{i\theta_1};\;\;\;\;
\phi_2=|\phi_2|e^{i\theta_2}
\ee
and then (\ref{bpsvortex}) become (after taking derivatives and making the combinations $f_{12}^{(i)}$)
\bea
  f_{12}^{(1)}&=&2\d\bar\d\ln |\phi_1|^2-2i(\d\bar\d-\bar\d \d)\theta_1=\frac{1}{2}\Delta \ln|\phi_1|^2+\epsilon^{ab}  
  \d_a\d_b \theta_1\,,\nonumber\\
  \\
  -f_{12}^{(2)}&=&2\d\bar\d \ln |\phi_2|^2+2i(\d\bar\d-\bar\d\d)\theta_2=\frac{1}{2}\Delta\ln|\phi_2|^2-\epsilon^{ab}  
  \d_a\d_b\theta_2\,.\nonumber
\eea
But since, if $\a$ is the polar angle in the $1,2$ plane, $\epsilon^{ab}\d_a\d_b \a=2\pi \delta^2(x)$
(as can be checked by integrating over a circle of vanishingly small radius), we may take the ansatz
\be
  \theta_1=-N_1\a\,,\;\;\;
  \theta_2=N_2\a\,,
\ee
which leads to 
\bea
  f_{12}^{(1)}&=&\frac{1}{2}\Delta \ln |\phi_1|^2-2\pi N_1\delta^2(x)\,,\nonumber\\
  \\
  -f_{12}^{(2)}&=&\frac{1}{2}\Delta \ln |\phi_2|^2-2\pi N_2\delta^2(x)\,.\nonumber
\eea
Finally, on substituting the Gauss law constraints, we obtain a continuous Toda system with delta functions sources,
\bea
  \Delta\ln |\phi_1|^2&=&-\left(\frac{4\pi}{k}\right)^2|\phi_2|^2\left(|\phi_1|^2+\frac{\mu k}{2\pi}\right)+4\pi   
  N_1\delta^2(x)\,,\nonumber\\
  \\
  \Delta\ln |\phi_2|^2&=&-\left(\frac{4\pi}{k}\right)^2|\phi_1|^2\left(|\phi_2|^2-\frac{\mu k}{2\pi}\right)+4\pi   
  N_2\delta^2(x)\,,\nonumber
  \label{toda}
\eea
whose solutions we proceed to analyze.

\subsection{Asymptotic Analysis of the Toda System}
As in the case of the (much simpler) Nielsen-Olesen vortex, the Toda system of equations does not, as far as we are aware, exhibit any closed form analytic solution. Consequently, here too must we resort to topological, asymptotic and numerical analyses to tease out finite energy solutions from it. The argument, fortunately, goes through in much the same way as for the Neilsen-Olesen case: the topological term in the energy
\be
  \int_{\mathbb{R}^{2}} dx dy\, \mu j_0=\int_{\mathbb{R}^{2}} dx dy\,\frac{\mu k}{2\pi}\,\Tr(F_{12})=
  \frac{\mu k}{2\pi}\frac{N(N-1)}{2}\int_{\mathbb{R}^{2}} dx dy\,\left(f_{12}^1+f_{12}^2\right)\,,\nonumber
\ee
is quantized as usual, since for an abelian gauge field
\be
  \frac{1}{2\pi}\int_{\mathbb R^2}F_{12}\,dx dy=\frac{1}{2\pi}\oint_CA_\a dl
  =\frac{1}{2\pi}\int_0^{2\pi}A_\a d\a\,,\nonumber
\ee
and $D_a\phi\rightarrow 0$ at $r\rightarrow\infty$, with $\phi=|\phi|e^{i\theta}$, $D_\mu=\d_\mu-ia_\mu$ and $\d_\mu \ln |\phi|\rightarrow 0$, means that $\d_a\theta^\infty -A_\a^\infty=0$ and consequently
\be
  \int_{\mathbb R^2}F_{12}dx dy=2\pi \tilde{N}\,.\nonumber
\ee
The corresponding statement for our system (\ref{bpsvortex}) is that, at infinity
\be
  \d_\a \theta^\infty_i=a_\a^{(i)\infty}\,,\nonumber
\ee
which gives the energy of the BPS state as
\be
  E(N_1,N_2)=\mu k \frac{N(N-1)}{2}(N_1+N_2)\,.
\ee
We are now in a position to look at the Toda equations in the asymptotic regions. To obtain the $r\rightarrow 0$ behaviour, we integrate each of them over a very small disk of radius $R\rightarrow 0$, and find that
\be
  \int dx dy\, \vec{\nabla}\cdot \vec{\nabla}\ln |\phi_{i}|=2\pi N_{i}\,,\quad i = 1,2\,,\nonumber
\ee
which, after using Stokes' theorem, gives
\be
  R\frac{d}{dr}\ln |\phi_{i}||_{r=R}=N_{i}\,.\nonumber
\ee
This expression is easily integrated to show that as $r\rightarrow 0$ each of the scalars exhibits the power law behaviour, 
\be
 |\phi_{i}|\sim A_{i} r^{N_{i}}\,.
\ee
This is in accordance with the usual argument says that the only possibility for the vortices with $\phi=|\phi|e^{iN\a}$ is to have $|\phi|\rightarrow 0$ at $r\rightarrow 0$ in order that the phase is well defined at $r=0$. 
In fact we can do better and refine the conditions at $r\rightarrow 0$ by using the equations of motion away from $r=0$. Taking as an ansatz for the scalars
\bea
  |\phi_1|^2&=&A_{1}r^{2N_1}(1+B_{1}r^p)\,,\nonumber\\
  \\
  |\phi_2|^2&=&A_{2}r^{2N_2}(1+B_{2}r^q)\,,\nonumber
\eea
and substituting into the Toda equations, we find
\bea
  &&q=2N_1+2,\;\; p=2N_2+2\,,\nonumber\\
  && B_{1}=\frac{8\pi\mu}{kp^2}A_{2}\,,\\
  && B_2=-\frac{8\pi \mu}{kq^2}A_1\,.\nonumber
\eea
The constants $A_{i}$ are only determined from the full numerical solution.

At $r\rightarrow \infty$, we can first check that neither a constant, nor a decaying exponential, nor a power law that blows up, $\phi\sim r^p$ works for either of the two fields. This leaves a decaying power law as the only plausible behaviour for either of the two scalar fields. In order to use the equations above, we must 
consider also the first subleading terms, {\it i.e.} we substitute 
\bea
  |\phi_1|^2&=&\frac{\bar A_1}{r^m}\left(1+\frac{\bar B_1}{r^p}\right)\,,\nonumber\\
  &{}&\\
  |\phi_2|^2&=&\frac{\bar A_2}{r^n}\left(1+\frac{\bar B_2}{r^q}\right)\,,\nonumber
\eea
into the equations above, to find
\bea
&&m=q+2,\;\; n=p+2\,,\nonumber\\
&&{\bar B}_{1}=-\left(\frac{8\pi \mu}{p^{2}k}\right){\bar A}_{2}\,,\\
&&{\bar B_2}=\left(\frac{8\pi \mu}{q^{2}k}\right){\bar A_{1}}\,.\nonumber
\eea
Since $p,q\in \mathbb N_*$, $m,n=3,4,5,...$ Again, the constants $\bar A_1,\bar A_2$, as well as $m,n$ are determined from the full numerical solutions. 
\subsection{Numerical Analysis of the Toda System}
To determine the various parameters of the vortex-like solutions described above, we need to solve the Toda system
numerically. As in the asymptotic analysis above, our numerical solution follows the general logic of the Abelian-Higgs model. Specifically, we will use a modified {\it two-parameter shooting method} to numerically solve the two-point boundary value problem described by the coupled Toda equations. To facilitate the implementation of the shooting algorithm, we first rewrite the equations as a four-dimensional (non-autonomous) dynamical system. To this end, we first non-dimensionalize the system by rescaling our variables and defining
\begin{eqnarray}
  g&\equiv& \frac{2\pi}{\mu k}|\phi_{1}|^{2},\nonumber\\
  f&\equiv& \frac{2\pi}{\mu k}|\phi_{2}|^{2},\\
  R&\equiv& \frac{r}{2\mu},\nonumber
\end{eqnarray}
with $r$ taken to be the radial coordinate on the plane. Substituting into the system (\ref{toda}) and assuming that the solitons that we are looking for are rotationally symmetric on the plane (so that the two-dimensional Laplacian is 
$ \Delta = \frac{1}{r}\frac{d}{dr}\left(r\frac{d}{dr}\right)$), we find that eqs.(\ref{toda}) reduce to
\begin{eqnarray}
  \frac{1}{R}\frac{d}{dR}\left(\frac{R}{f}\frac{df}{dR}\right) = -g\left(f-1\right),\nonumber\\
  {}\\
  \frac{1}{R}\frac{d}{dR}\left(\frac{R}{g}\frac{dg}{dR}\right) = -f\left(g+1\right).\nonumber
\end{eqnarray}
Finally, we reduce the order of the system by one by making the additional definitions $h\equiv\frac{df}{dR}$ and 
$j\equiv\frac{dg}{dR}$ so that (denoting by a $'$ derivatives with respect to the dimensionless radial variable $R$)
\begin{eqnarray}
  f' &=& h,\nonumber\\
  h' &=& \frac{h^{2}}{f} - \frac{h}{r} - gf(f-1),\nonumber\\
  g' &=& j,\\
  j' &=& \frac{j^{2}}{g} - \frac{j}{r} - gf(g+1).\nonumber
\end{eqnarray}
Before directly integrating this four-dimensional non-autonomous dynamical system, it will be instructive to extract some qualitative information from it. There are two (physical) fixed points at $(f,h,g,j) = (0,0,0,0)$ and $(1,0,0,0)$. A linearization of the system near the former, shows that the origin is a saddle. Solutions of the kind that carry nonvanishing winding number and conform to the asymptotic boundary conditions $f(0)=g(0)=0$ and $f(\infty)=g(\infty)=0$ correspond to {\it homoclinic} orbits\footnote{This should be compared to the standard ANO vortices of the Abelian-Higgs model which correspond to {\it heteroclinic} orbits interpolating between the two fixed points of the associated dynamical system.} that begin and end at $(0,0,0,0)$ and that encircle the fixed point at $(1,0,0,0)$ (see Fig.1).

\begin{figure}[h]
  \centering
  \includegraphics[width=4cm,height=4cm]{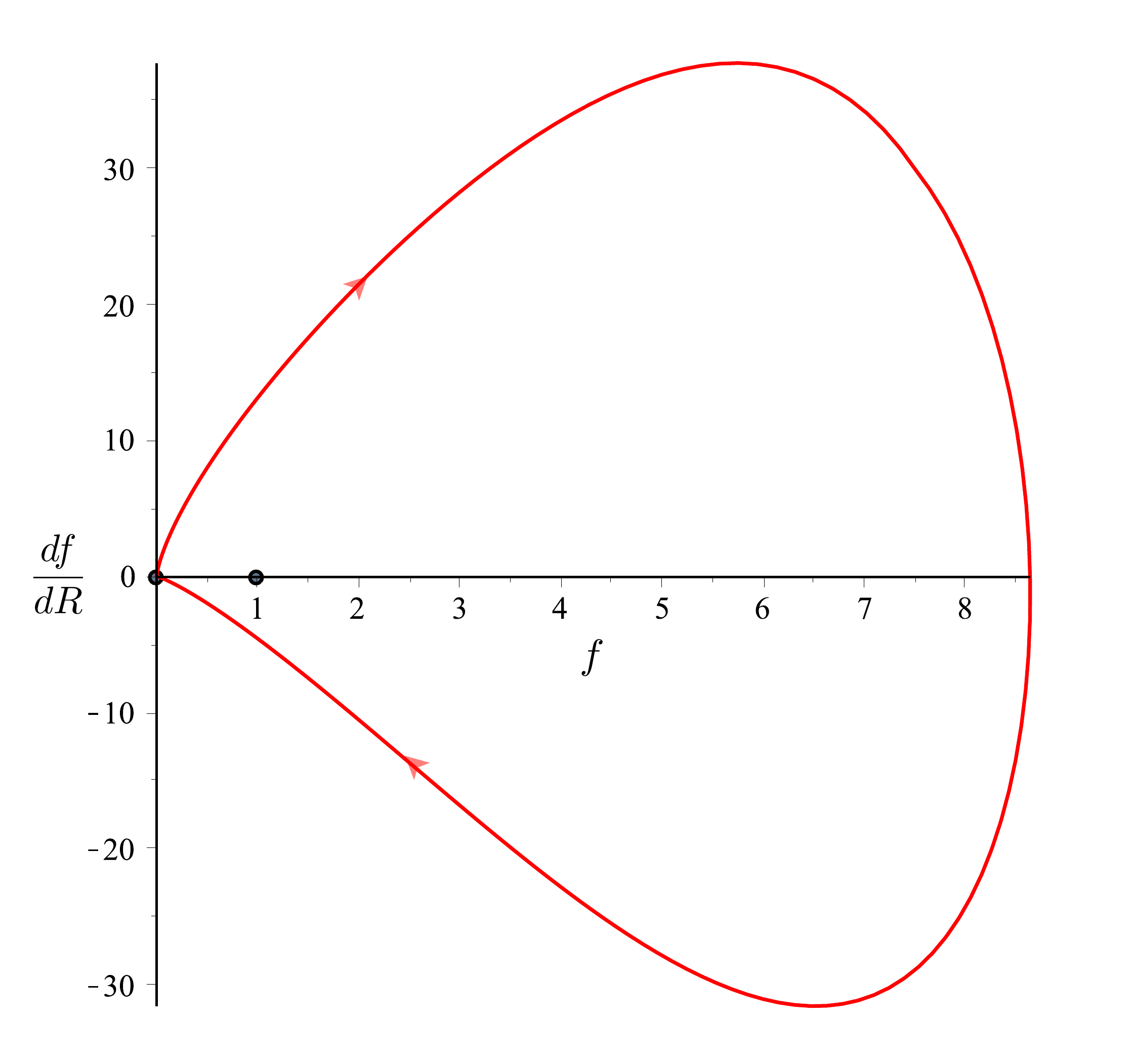}
  \caption{The $(f,f')-$subspace of the full phase space of the Toda system}
\end{figure}  

Our numerical integration of the system is based on a two-parameter shooting algorithm that converts the nonlinear dynamical system above into a nonlinear parameter estimation problem. The parameters in question are precisely the undetermined constants $A_{1}$ and $A_{2}$ above and these are chosen at $R=0$ so that the {\it constraint} $f(\infty)=g(\infty)=0$ is met. In practice, the constraints at $R=\infty$ are a problem, but our asymptotic anaylsis above can be extended to show that solutions at $R\approx10$ are quite safely in the far field for both $f$ and $g$. Some results of our numerical integration are presented in Figures 2 and 3. 
\begin{figure}[h]
    \centering
    \includegraphics[width=5cm,height=5cm]{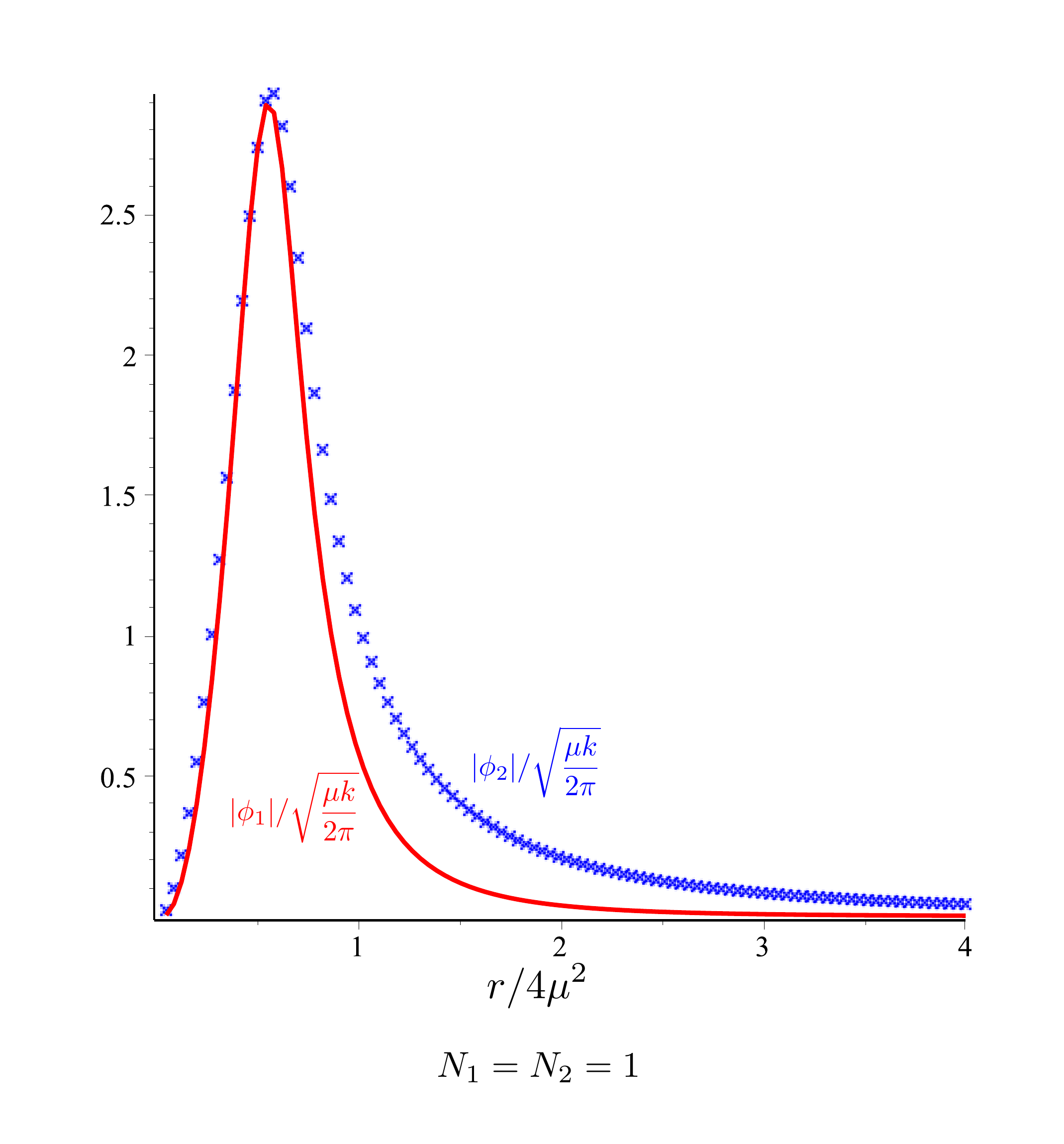}
    \caption{The $N_{1}=N_{2}=1$ soliton profiles. Optimization of the shooting parameters yeild 
    $A_{1}=30.00, A_{2}=30.05$}
\end{figure}     
  
\begin{figure}[h]
    \centering
    \includegraphics[width=5cm,height=5cm]{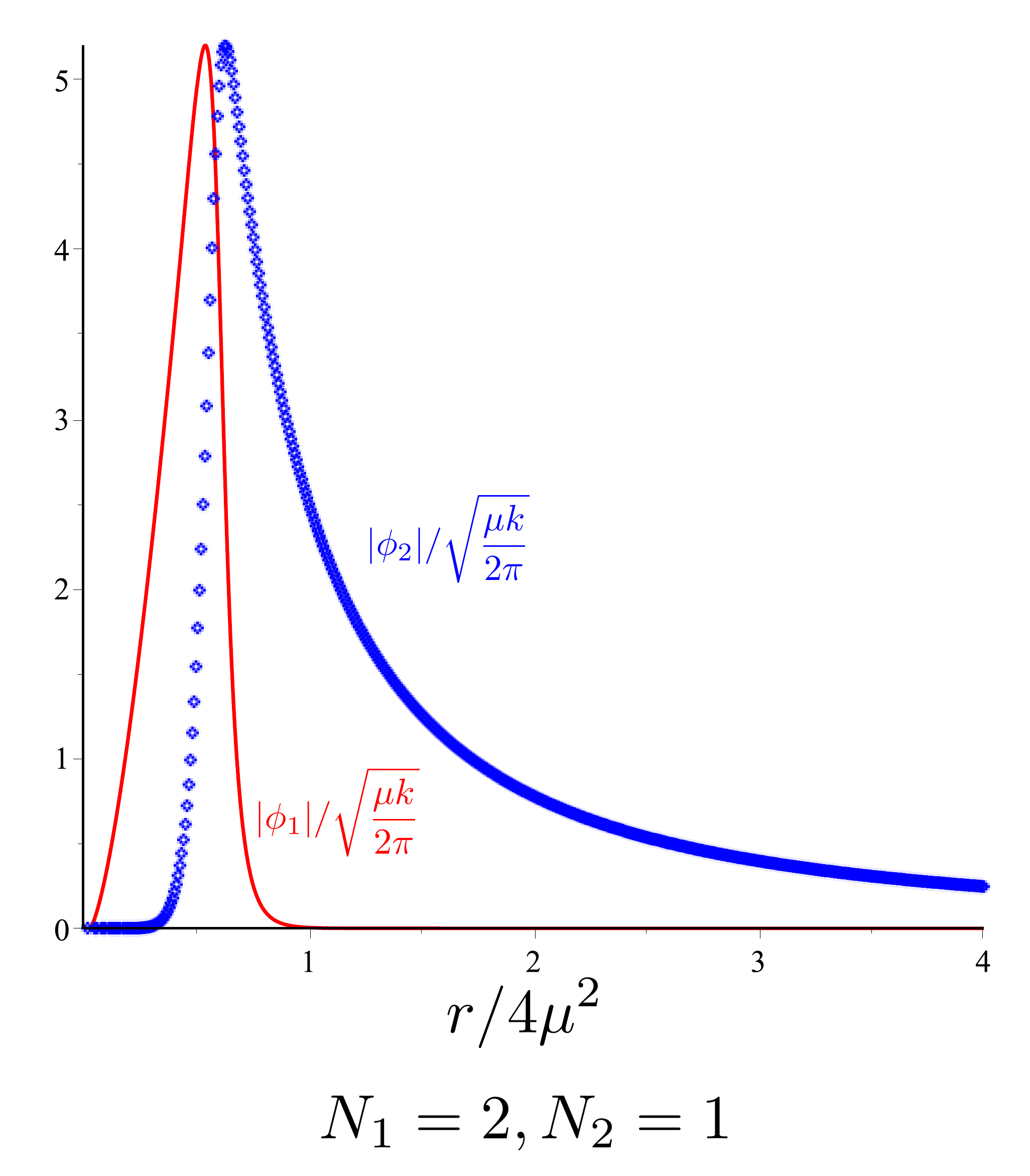}
    \caption{The $N_{1}=2, N_{2}=1$ soliton profiles. Optimization of the shooting parameters yeild 
    $A_{1}=50.00, A_{2}=100.00$}
  \end{figure}     

We also obtain from the numerics that the power law at infinity is $|\phi_1|^2\propto 1/r^3, |\phi_2|^2\propto 1/r^2$,
i.e. $m=3,n=2$.

As a final point, we note that in the massless ABJM case, at $\mu=0$, these vortices vanish since their energy is proportional to $\mu$. This agrees well with known facts about the solitonic spectrum of pure ABJM \cite{Arai:2008kv}.
\section{The Abelian-Higgs model from ABJM}
We now look to embed the Abelian-Higgs model in ABJM, as a truncation of our general abelianization ansatz. To find the truncation we look at the multi-vortex solution we found previously in \cite{Mohammed:2010eb} 
for the $N=2$ case, i.e. $U(2)\times U(2)$ ABJM. There, not only was the ansatz written in a manner similar to 
the multi-vortices of the conventional Abelian-Higgs model, but the action on the moduli space of vortices was also found to be the same. In retrospect, this was really a telling signal that we were actually embedding the Abelian-Higgs model into ABJM. For the reader unfamiliar with \cite{Mohammed:2010eb}, we recall that the static multivortex solution there was given by
\bea
C^1&=&\sqrt{\frac{k\mu}{\pi}}\begin{pmatrix} 0&0\\0&1\end{pmatrix}=\sqrt{\frac{k\mu}{\pi}}G^1\,,\nonumber\\
C^2&=&\sqrt{\frac{k\mu}{\pi}}e^{-\psi/2}H_0(z)\begin{pmatrix} 0&1\\0&0\end{pmatrix}=\sqrt{\frac{k\mu}{\pi}}e^{-\psi/2}H_0(z) G^2;\;\;\;
C^3=C^4=0\,,\\
A_0&=&\frac{1}{\mu}\begin{pmatrix} \d \bar\d \psi & 0\\0&0\end{pmatrix};\;\;\;
\hat A_0=\frac{1}{\mu}\begin{pmatrix} 0&0\\0&\d\bar\d \psi\end{pmatrix};\;\;\;
A_{\bar z}=\hat A_{\bar z}=\begin{pmatrix} 0&0\\0&\frac{i}{2}\bar\d \psi\end{pmatrix}\,,\nonumber
\eea
where $H_0(z)=\prod_{i=1}^n(z-z_{\tilde i})$ is an arbitrary polynomial and the real function $\psi(z)$ is determined through the equation
\be
  \d\bar\d\psi=\mu^2\left(1-e^{-\psi}|H_0(z)|^2\right)
\ee
with boundary conditions at $|z|\rightarrow\infty$ requiring $\psi\rightarrow\log|H_0(z)|^2$. As usual, $z_i$ with $i=1\ldots n$, denotes the positions of the $n$ vortices. Treating each of these position variables as (adiabatic) functions of time, $z_{\tilde i}(t)$,  produces the first order solution
\bea
  C^I&=&0\,,\nonumber\\
  A_0^{(1)}&=&\hat A_0^{(1)}=\begin{pmatrix}0&0\\0&-\frac{i}{2}(\dot z^{\tilde i}\d_{\tilde i}-
  \dot{ {\bar z}}^{\tilde i} \bar \d_{\tilde i})\psi\end{pmatrix}\,,\\
  A_{\bar z}^{(1)}&=&\begin{pmatrix} \frac{1}{2\mu}\dot z^{\tilde i} \d_{\tilde i} \d \psi&0\\0&0\end{pmatrix}
  \,,\quad
  \hat A_{\bar z}^{(1)}=\begin{pmatrix} 0&0\\0&\frac{1}{2\mu}\dot z^{\tilde i} \d_i \d\psi\end{pmatrix}\,,  
  \nonumber
\eea
on the moduli space of the vortices. On the other hand, when $N=2$, our general abelianization ansatz gives
\bea
  A_\mu&=&a_\mu^{(1)} G^1G^\dagger_1+a_\mu^{(2)} G^2G^\dagger_2=\begin{pmatrix} a_\mu^{(1)}&0\\0&a_  
  \mu^{(2)}\end{pmatrix}\,,\nonumber\\
  {}\\
  \hat A_\mu&=&a_\mu^{(1)}G^\dagger_1G^1+a_\mu^{(2)} G^\dagger_2G^2=\begin{pmatrix}0&0\\0&a_\mu^{(1)}+a_  
  \mu^{(2)}\end{pmatrix}\,.\nonumber
\eea
Comparing with the solution above (and also denoting now the first order solution for the abelian fields with a 
tilde to avoid confusion with the indices $(1)$ and $(2)$ on the $a$'s) we find
\bea
  &&a_{\bar z}^{(2)}=\frac{i}{2}\bar\d \psi,\;\;\;
  a_0^{(1)}=\frac{1}{\mu }\d\bar\d\psi,\;\;\;
  a_a^{(1)}=a_0^{(2)}=0\,,\nonumber\\
  &&\tilde a_{\bar z}^{(1)}=\frac{1}{2\mu}\dot z^{\tilde i}\d_{\tilde i}\d\psi\,,\label{abarz}\\
  && \tilde a_0^{(2)}=-\frac{i}{2}(\dot z^{\tilde i}\d_{\tilde i}-\dot{\bar z}^{\tilde i}
  \bar\d_{\tilde i})\psi\,.\nonumber
\eea
Note that from (\ref{abarz}), by taking complex conjugate and then sums and differences, we get
\bea
  \tilde a_1^{(1)}&=&\frac{1}{2\mu}\left[\left(\dot{z}^{\tilde i}\d_{\tilde i}+\dot{\bar{z}}^{\tilde i}
  \bar{\d}_{\tilde i}\right)\d_1-i\left(\dot z^{\tilde i}\d_{\tilde i}
  -\dot{\bar{z}}^{\tilde i}\bar{\d}_{\tilde i}\right)\d_2\right]\psi\,,\nonumber\\
  \\
  \tilde a_2^{(1)}&=&-\frac{i}{2\mu}\left[\left(\dot z^{\tilde i}\d_{\tilde i}-\dot{\bar{z}}^{\tilde i}
  \bar{\d}_{\tilde i}\right){\d}_1-i\left(\dot{z}^{\tilde i}\d_{\tilde i}
  +\dot{\bar{z}}^{\tilde i}\bar{\d}_{\tilde i}\right)\d_2\right]\psi\,.\nonumber
\eea
This can be written more compactly, by using the fact that $\psi$ is a real-valued field, as
\be
  \tilde a_i^{(1)}=\frac{i}{2\mu}\epsilon_{ij}\left(\dot z^{\tilde k}\d_{\tilde k}-\dot{\bar{z}}^{\tilde k}
  \bar\d_{\tilde k}\right)\d_j\psi\,.
\ee
In view of the above solution, and assuming that the same relation to our abelianization holds at all $N$, 
we can now identify the truncation ansatz needed to obtain the abelian-Higgs model as 
\be
  \phi_1=\phi_2=0,\chi_1=b={\rm constant}\,,
\ee
which gives
\bea
  &&D_\mu \phi_1=D_\mu \phi_2=0\,,\nonumber\\
  &&D_\mu \chi_1=-ia_\mu^{(1)}b\\
  &&D_\mu \chi_2=(D_\mu -ia_\mu ^{(2)})\chi_2\nonumber
\eea
and the potential
\bea
V&=&\frac{2\pi^2}{k^2}N(N-1)[|b|^2(|\chi_2|^2-c^2)^2+|\chi_2|^2(|b|^2-|c|^2)^2]\,,\nonumber\\
&=&\frac{2\pi^2}{k^2}N(N-1)[|b|^2|\chi_2|^4+|\chi_2|^2(-4|b|^2c^2+|b|^4+c^4)+c^4|b|^2]\,.\label{abhigpot}
\eea
We can easily arrange for the coefficient of the $|\chi|^2$ term to be negative, as is required for the mexican hat potential of the abelian-Higgs model, by choosing for instance
\be
  |b|=|c|\Rightarrow \mu=\frac{2\pi |b|^2}{k}\,.
\ee
The action is then
\be
  S=-\frac{N(N-1)}{2}\int d^3x\left[\frac{k}{2\pi}\epsilon^{\mu\nu\lambda}a_\mu^{(1)}f_{\nu\lambda}^{(2)}+  
  \left(a_\mu^{(1)}\right)^2|b|^2+|D_\mu \chi_2|^2+V\right]\,,
\ee
with the auxiliary field $a_\mu^{(1)}$. As usual it can be eliminated through its equation of motion
\be
  a_\mu^{(1)}=-\frac{k}{4\pi |b|^2}\epsilon^{\mu\nu\lambda}f_{\nu \lambda}^{(2)}\,.\label{af}
\ee
so that 
\be
  S=-\frac{N(N-1)}{2}\int d^3x\left[\frac{k^2}{8\pi^2|b|^2}\left(f_{\mu\nu}^{(2)}\right)^2+|D_\mu \chi_2|^2
  +V\right]\,,\label{abhiggs}
\ee
which is nothing but the action of the abelian-Higgs model.

Of course, we still need to check the consistency of the truncation, i.e. to check that the equations of motion of the full abelianization ansatz in section 2 are satisfied. We have fixed $\phi_1,\phi_2$ to zero and $\chi_1$ to $b$, so it is these three equations of motion that we need to check. As before, 
the choice $\phi_1 = \phi_2 =0$ is a consistent truncation. The equation for $\chi_1$ reduces, 
in the Lorentz gauge $\d^\mu a_\mu^{(2)}=0$, and using (\ref{af}), to 
\be
  \left(a_\mu^{(1)}\right)^{2}b =b\frac{\d V}{\d |b|^2}
\ee
which is just the equation of motion we would obtain for the parameter $b$ by varying in the abelian-Higgs action (\ref{abhiggs}). We find this somewhat puzzling, since it means that the {\it constant parameter} $|b|$ has to be effectively treated like a field in the abelian-Higgs action, giving its own equation of motion.

It remains now to check that our multivortex solution satisfies the condition (\ref{af}), since it certainly matched our ansatz {\em before} we imposed the equation of motion for $a_\mu^{(1)}$. The equations (\ref{af}) reduce for the zeroth order solution and the first order solution
respectively, to 
\bea
  a_0^{(1)}&=& \frac{k}{4\pi |b|^2}\epsilon^{ij}\d_ia_j^{(2)}\nonumber\\
  \\
  \tilde a_i^{(1)}&=&-\frac{k}{4\pi|b|^2}\epsilon^{ij}\d_j \tilde a_0^{(2)}\nonumber
\eea
We can check the first equation, since $a_{\bar z}^{(2)}=\frac{i}{2}\bar \d \psi$, which written in real components reads $\tilde a_i^{(2)}=-\frac{1}{2}\epsilon_{ij}\d_j\psi$. Then the zeroth order equation is satisfied if
\be
\mu=\frac{2\pi |b|^2}{k}
\ee
while the first order equation is satisfied when 
\be
\tilde a_i^{(1)}=\frac{i}{2\mu}\epsilon_{ij}(\dot z^{\tilde k}\d_{\tilde k}-\dot \bar z^{\tilde k}\bar\d_{\tilde k})\d_j\psi
\ee
as expected. The appearance of the abelian-Higgs model above is somewhat non-standard but can easily be put into canonical form by appropriately normalizing the fields as
\[
  a^{(2)}=\frac{2\pi b}{Nk}\tilde a^{(2)}\,,\quad \chi_2=\frac{\tilde \chi_2}{N}\,,
\]
to obtain 
\be
  S=\int d^3x\left[-\frac{1}{4}\left(\tilde f^{(2)}_{\mu\nu}\right)^2-|D_\mu\tilde \chi_2|^2-V\right]
\ee
where now $D_\mu=\d_\mu-ig\tilde a^{(2)}_\mu$ and $g=\frac{2\pi |b|}{Nk}$. In terms of the canonical fields and coupling, the potential 
\be
  V=\frac{g^2}{2}\left[|\tilde \chi_2|^4+\frac{\mu^2k^2N^4}{4\pi^2}+|\tilde \chi_2|^2N^2
  \left(-\frac{4\mu k}{2\pi}+|b|^2 +\frac{\mu^2k^2}{4\pi^2 |b|^2}\right)\right]\,.\label{potencanon}
\ee
As previously alluded to, the potential has a range of values of $|b|$ for which it is spontaneously breaking (has negative mass squared). The central value of this
domain is $|b|=c$, and for this value of $|b|$, we find
\be
  V=\frac{g^2}{2}\left[|\tilde \chi_2|^2-\frac{\mu k N^2}{2\pi}\right]\,.
\ee
Moreover, for this value of $|b|$, the equation of motion for $|b|^2$  (the extra constraint on our abelian-Higgs model), becomes 
\be
  \left[|\phi_2|^2-c^2\right]^2=\frac{1}{2}\left[\frac{k}{2\pi \mu}f_{\mu\nu}^{(2)}\right]^2\,.
\ee
On the other hand, equating the kinetic (Maxwell) term for $f_{\mu\nu}^{(2)}$ with the potential 
term, $V$, gives exactly the same equation. Further, taking the square root of this equation, and imposing that$f_{0i}=0$, we find
\be
  \frac{1}{Ng}f_{12}^{(2)}=\pm[|\tilde \chi_2|-N^2c^2]\,,
\ee
which is part of the abelian-Higgs BPS condition. In other words, the extra condition is satisfied on BPS solutions of the abelian-Higgs model with $f_{0i}=0$, and in particular for vortices.

Having established the consistent truncation to the Landau-Ginzburg model of interest, as explained in the introduction, we still need to establish the conditions under which we can decouple the nonzero modes. From the potential (\ref{potencanon}) we find that generically the mass term has $m\sim \mu$ (for instance for 
the central value $|b|^2=\mu k/(2\pi)$), whereas it vanishes at 
\be
|b|^2=\frac{\mu k}{2\pi}(2\pm \sqrt{3}), 
\ee
so, for values close to these, the mass of $\tilde\chi_2$ can be made much smaller than $\mu$. The coupling $g^2$ is generically, for instance close to the 
central value of $|b|^2$, 
\be 
g^2=\frac{4\pi^2|b|^2}{N^2k^2}\sim \mu\frac{2\pi}{N^2k}.
\ee
If we choose large $N$ (as is required for the gravity dual) and $k\sim 1$, we see that $g^2\ll \mu$. The constant term is large in that case, for instance for 
near massless $\tilde\chi_2$, 
\be
\frac{1}{2}\frac{g^2\mu^2 k^2N^4}{4\pi^2}\simeq\frac{\mu kN^2}{4\pi}(2\pm \sqrt{3}).
\ee
Since we are in a non-gravitational theory here, this cannot be measured and, consequently, it does not matter.

It is also possible to analyze the various terms in the ABJM action to see which of them are quadratic in the nonzero modes since, these will be the terms responsible for the simplest quantum loops. We find, using the ansatz for the ``zero mode" fields that give the LG action, and considering the nonvanishing
$\delta\phi$ as generic $(ij)$ modes in the $N\times N$ matrices, 
\bea
&&\frac{4\pi^2}{3k^2}\Tr[C^6]\sim \frac{1}{k^2N^2}\tilde\chi_2^4(\delta\phi)^2\propto\frac{1}{N^2},\cr
&&\frac{k}{4\pi}\Tr[A^3]\sim kN a (\delta\phi)^2\propto N,\cr
&&\Tr[D_\mu C^ID^\mu C_I]\sim (\d_\mu \delta\phi)^2;\;\;\;\; \Tr[\psi^\dagger \Dslash\psi]\sim \delta\psi^\dagger\dslash\delta\psi,\cr
&&\frac{2\pi}{k}\Tr[C^\dagger C\psi^\dagger\psi]\sim\frac{1}{kN}(\tilde\chi_2)^2\delta\psi^\dagger\delta\psi\propto\frac{1}{N},\cr
&&\frac{2\pi\mu}{k}\Tr[Q^\dagger Q R^\dagger R]\sim \frac{\mu}{kN}(\tilde\chi_2)^2(\delta\phi)^2\propto \frac{\mu}{N}.
\eea
Evidently then, the Chern-Simons term generates a term with large coupling, and the mass term generates a term $\propto \mu/ N\ll \mu$, but still $\gg \mu/N^2\sim g^2$. These couplings cannot be made small; leading us  to the situation that we advertised: the masses of the nonzero modes are much larger than the mass parameters of the reduced theory, while the couplings remain relatively large. 

Nevertheless, we still need to show that the modes of the reduced theory are the only light ones in the theory or, if they are not, 
that any additional light modes do not couple to ours. To this end, let's start with the other modes in (\ref{abelianmaster}). Rescaling to canonically normalized fields, 
\be
\chi_i=\frac{\tilde\chi_i}{N};\;\;\;\phi_i=\frac{\tilde\phi_i}{N};\;\;\; 
a^{(2)}=\frac{2\pi b}{Nk}\tilde a^{(2)},
\ee
we find the following in the absence of a Higgs VEV: 
\begin{itemize}
  \item
    Sextic terms in the scalars go like $\sim1/N^4k^2\rightarrow 0$, 
  \item  
    Quartic terms go like $\sim c^2/N^2k^2\sim \mu/(N^2g)\sim g^2$, and 
  \item  
    Mass terms go like $\sim \mu^2$. 
\end{itemize}    
As claimed, they are generically heavy. All that remains then is to check what happens in the presence of the Higgs VEV $\chi_1=b$. 
In this case we obtain the extra terms 
\be
  \frac{1}{k^2}[-4b^2c^2|\tilde\chi_2|^2+b^4(|\tilde\chi_2|^2+|\tilde\phi_2|^2)],
\ee
so that only the $\tilde\chi_2$ mode can become light; all others remain massive. What about
generic modes outside the action (\ref{abelianmaster})? We already saw that generic mass terms are of order
$m^2\sim\mu^2>0$, so it only remains to see that the terms coming from the Higgs VEV $\chi_1=b$ cannot 
cancel them. Thus we search for solutions to the vanishing of the 
mass term coming from the ABJM action, where we only keep
two $C^I$'s general in each term, and the rest we write as 
\be
C^I=(R^1=bG^1,R^2=Q^1=Q^2=0).
\ee
Setting this mass term to zero produces a very long equation, for the trace of a sum of 
terms with two $C^I$ matrices and up to four $G^1$ matrices being 0. One solution of this equation is given by our light mode
\be
C^I=(R^2=\chi_2G^2,R^1=Q^1=Q^2=0);\;\;\; b^2=c^2(2\pm \sqrt{3}),
\ee
and is equivalent to an identity between $G^1$ and $G^2$ matrices after the ansatz has been considered. 
The issue is whether or not the solution is unique. While we don't know a mathematical proof of uniqueness, physically 
it is clear it should be so. Indeed, the solution is related to the existence of the maximally supersymmetric
fuzzy sphere vacuum characterized by $G^1,G^2$; once we turn on $G^1$, there is an instability towards turning
on $G^2$ as well, apparent in the fact that the mass of $\chi_2$ can go through zero and become negative.
Any other solution would amount to the statement that there is another vacuum with $G^1$ turned on (corresponding to 
a different instability in the presence of $R^1=bG^1$). As there is no other vacuum connected in this way 
to the maximally supersymmetric one, we conclude that there should be no other solution to the zero mass 
equation. Hence there are no other light modes in the presence of the Higgs VEV $\chi_1=b$. Of course, there 
can be other light modes in other regions of parameter space, but all we need is that for large $N$, $k\sim 1$ 
and the only VEV turned on being $R^1=bG^1$ we don't have other light modes, and we have argued this is indeed the 
case.

This completes our demonstration that (i) the abelian Higgs model can be obtained from the abelianization of the ABJM model as a quantum consistent truncation and (ii) that both the classical zeroth order and the first order (in the moduli space approximation)  multivortex solutions of the latter at $N=2$, are encoded in this model. Since this is a {\it bone fide} embedding of the abelian-Higgs model, we say say more even. For instance, it is natural that we obtain the same
fluctuation action for vortex scattering as in the abelian Higgs case. It also means that we can now immediately 
write down the multivortex solution at general $N$, with the guarantee that we will recover the same fluctuation 
action for vortex scattering as in the abelian Higgs case. To be concrete, the multivortex solution at general $N$ in ABJM 
\bea
  R^1&=&\sqrt{\frac{k\mu}{\pi}}G^1,\;\;\;
  R^2=\sqrt{\frac{k\mu}{\pi}}e^{-\psi/2}H_0(z) G^2,\;\;\;
  Q^1=Q^2=0\,,\nonumber\\
  A_0&=&\frac{1}{\mu}\d\bar\d\psi G^1G^\dagger_1,\;\;\;
  \hat A_0=\frac{1}{\mu}\d\bar\d\psi G^\dagger_1 G^1,\\
  A_{\bar z}&=&\frac{i}{2}\bar\d\psi G^2 G^\dagger_2,\;\;\;
  \hat A_{\bar z}=\frac{i}{2}\bar\d \psi G^\dagger _2 G^2\,,\nonumber
\eea
produces an effective Lagrangian on the moduli space  
\bea
L_{eff}&=&\frac{N(N-1)}{2}\frac{k\mu}{\pi}\int d^2x \left[-\d\bar\d\psi+\frac{1}{2}\dot z^i\dot{\bar{z}}^j
\left(\d_i\bar\d_j \psi+\frac{1}{\mu^2}(\d_i\bar\d\psi\bar\d_j\d\psi-\d\bar\d\psi\d_i\bar\d_j\psi)\right)\right]\,,
\nonumber\\
&\simeq& \frac{N(N-1)}{2}\left[
-k \mu n+\sum_{i=1}^n\frac{k\mu}{2}|\dot{z}^i|^2-k\mu q \sum_{i>j}K_0(2\mu|z^i-z^j|)|\dot{z}^i-\dot{z}^j|^2\right]\,,
\eea
with $q\simeq 1.71$. To close this discussion on vortices of the abelian-Higgs model and their embedding into the ABJM model, we mention briefly that in the case of the massless ABJM model, with $c=\mu=0$, we obtain a non-symmetry breaking potential, 
\be
  V=\frac{2\pi^2}{k^2}N(N-1)\left[|b|^2|\chi_2|^4+|b|^4|\chi_2|^2\right]\,,
\ee
which is just a massive gauged $\phi^{4}$ model.

\section{Towards a string construction of AdS/CMT}

At this point, let's stop and consider what it is that we have achieved. Stripping away all the bells and whistles, 
essentially our truncation has produced a (2+1)-dimensional scalar field theory with potential
\bea
  V =\frac{2\pi^2}{k^2}N(N-1)\left[|b|^2|\phi|^4+|\phi|^2((|b|^2-c^2)^2-2c^2|b|^2)+c^4|b|^2\right]\,.
  \label{lgpot}
\eea
It is not too difficult to see that it is just a {\it Landau-Ginzburg model} in which, at fixed $|b|^2$, $c^2\propto \mu$ acts as a coupling that takes us from a $|\phi|^4$ theory
(the insulator phase) to an abelian-Higgs theory (the superconducting phase). In this sense, the parameters 
$|b|^2$ and $c^2$ control the coupling $g$ and critical coupling $g_c$ of 
the Landau-Ginzburg model. More precisely, we identify the combinations $(|b|^2-c^2)^2$ as $g$ and $2c^2|b|^2$ as $g_c$ respectively. In this light, it makes sense then to think of this abelianization as {\it a realization of the recently proposed AdS/CMT correspondence}. To see why our construction is markedly different from any of its pre-cursors, we recall the general ideas involved. Usually, in an AdS/CMT construction, one assumes some theory in an AdS background, usually involving gravity, a gauge field $A_\mu$, maybe a complex (charged)
scalar $\phi$ and some fermions $\psi_i$. It is then argued that this theory 
should be dual to some large $N$ conformal field theory with a global current $J_\mu$ dual to the gauge field $A_\mu$, and some other operators (in principle) dual to the other fields. 
It is then argued that relevant physics in AdS corresponds to some behaviour of the operators in the field theory which simulates the relevant physics, like superconductivity \cite{Gubser:2008px} for example, to be studied. 
Sometimes the AdS theory is obtained as a consistent truncation of some known AdS/CFT duality 
(for which there is a heuristic derivation involving a decoupling limit of some brane constructions), 
so that the field theory contains a small subset of operators that could possibly give the desired physics \cite{Hartnoll:2009sz,Herzog:2009xv}.

However, even in these cases, it is not obvious how to directly relate the set of operators in the given 
CFT to the condensed matter system of interest, and usually one has to invoke some sort of universality argument. In other words, if the physics of the 
selected set of operators in the large $N$ CFT describes
the correct physics for the condensed matter system, then perhaps the physics is general 
enough to appear in many different systems, and we can 
try to apply our seemingly unrelated field theory to the condensed matter system of interest. While we certainly appreciate the logic of this argument, we find it less than satisfactory for a number of reasons. Primary among these is that it is not at all clear why can we choose only a very small number of operators in the large $N$ CFT and concentrate on their physics. Secondly, if we try to write 
down a gravity dual of an abelian theory having this small number of nontrivial operators, 
we would fail, since the absence of the large $N$ would mean that we could not focus on the supergravity limit in the dual.

However, we can now do better. We have found a {\em consistent truncation} of the 
large $N$ CFT, for which there is a well-defined duality, 
and not just a truncation of the gravity theory. That means that this set of fields is a 
well defined subset at the {\it quantum level}\footnote{We can 
consistently put the other fields to zero even at the quantum level.} corresponding to the collective 
motion of the nonabelian fields in the large $N$ case and involving ${\cal O}(N)$ out of the 
${\cal O}(N^2)$ fields of ABJM, via the nontrivial
matrices $G^\a$ (which have ${\cal O}(N)$ nonzero elements). It is not just a 
simple restriction to $N=1$ of the ABJM model, which would 
imply losing the supergravity limit in the dual. Therefore this abelianization 
still maps to a purely gravitational theory, and not a full string
theory as for generic abelian theories. 

We should note that the potential (\ref{abhigpot}) for $|b|=c$ has a minimum (vacuum) at 
$\chi_2=|b|=c$, which is nothing but the fuzzy sphere vacuum of the massive ABJM model, 
and hence classical solutions of the reduced theory (LG) can be understood as some type of 
deformations of the fuzzy sphere. We will see other examples of similar classical solutions in the 
next section. Therefore all of these solutions, representing a collective motion of ${\cal O}(N)$ 
fields, correspond to finite deformations of the gravity dual, unlike any solutions that only turn on 
one mode. In this sense, as already explained, the property of classical gravity dual related to large $N$
is still preserved by our abelianization. 

In our case, there already exists a well defined gravity dual of the field theory. In the 
case of massless ABJM, that theory corresponds to M2-branes moving 
in the space $\mathbb R^{2,1}\times \mathbb C^4/\mathbb Z_k$, and the gravity dual 
({\it i.e.} the near-horizon limit of the backreacted background) is 
$AdS_4\times\mathbb{CP}^3$. In the case of the massive ABJM, the theory corresponds 
to M2-branes moving in a space defined in \cite{Mohammed:2010eb,Lambert:2011eg}
with the gravity dual described in \cite{Auzzi:2009es,Mohammed:2010eb}. 
Of course, we still would need to understand to what the truncation to 
$\langle\chi_1\rangle=b$ and $\chi_2\neq 0$
corresponds in this gravity dual in order to complete the picture, but we leave this for further work. 

Actually, as it turns out, the theory we obtain in the abelianization is also the relevant effective theory for 
a CMT construction. Indeed, as reviewed for instance in \cite{Sachdev:2011wg}, starting from the 
Hubbard model for spinless bosons hopping on a lattice of sites $i$ with short range repulsive interactions, 
\be
  H_b=-w\sum_{\langle ij\rangle}\left(b_i^\dagger b_j+b_j^\dagger b_i\right
  )+\frac{U}{2}\sum_i n_i(n_i-1)-\mu \sum_i n_i\,,
\ee
where $n_i=b_i^\dagger b_i$ and $w$ is the hopping matrix between nearest-neighbour sites, one 
obtains the relativistic Landau-Ginzburg theory
\be
  S=\int d^3x \left(-|\d_t\phi|^2+v^2|\vec{\nabla}\phi|^2+(g-g_c)|\phi|^2+u|\phi|^2\right)\,.
  \label{lg}
\ee
The effective field $\phi$ is obtained as follows. The ground state contains an equal number of 
bosons at each site, with the creation operators $a_i^\dagger$ producing extra particles at each site, 
and creation operators $h_i^\dagger$ that produce extra ``holes" at each site; ``antiparticles"
in the QFT picture. Then, as is usual in field theory, $\phi_i\sim \a_i a_i+\b_i h_i^\dagger$ 
is a discretized version of the complex field describing both particles and antiparticles, 
where $\a_i,\b_i$ are wavefunctions for the modes.

For $g<g_c$ we have an abelian-Higgs system, i.e. a superconducting phase, while for $g>g_c$ 
we have an insulator phase. The marginal case $g=g_c$ is a conformal field theory. 
The systems described by the above model also have a quantum critical 
phase which opens up at nonzero temperature for a 
$T$-dependent window around $g=g_c$. This quantum critical phase is strongly coupled and very 
hard to describe using conventional condensed matter methods, 
which makes it an excellent choice for a holographic description. In \cite{Myers:2010pk}  
it was shown that by considering a gauge field in the gravity dual of ABJM and
introducing a coupling for it to the Weyl curvature, $\gamma \int C_{abcd}F^{ab}F^{cd}$ 
one obtains a conductivity $\sigma(\omega)$ consistent 
with the quantum critical phase, and from which it was concluded that ABJM is a good primer for 
these systems, though the precise reason for the match was not obvious.

While the bosonic Hubbard model leads, in the continuum limit to the action (\ref{lg}), the model itself is a 
drastic simplification, of a condensed matter system. The model has been used to describe the quantum critical phase of (bosonic) ${}^{87}Rb$ cold atoms on an optical lattice, but the description is believed to hold more generally for
the quantum critical phase. For instance, high $T_c$ superconductors have a ``strange metal" phase that is believed 
to be of the same quantum critical type. We can consider a solid with free electrons 
(fermions, perhaps several per atom) that could hop between fixed atoms, and unlike the simple Hubbard model, 
we also have in principle interactions that are not restricted 
to nearest neighbours. One could, for instance, generate bosons $\phi_{ij}$ (having the role of the bosons $b_i$ 
of the Hubbard model) by coupling fermions at two sites $i$ and $j$. By an abuse of notation we will call by the same $\phi_{ij}$ the field 
obtained by multiplying the corresponding ``particle creation" operator with a wavefunction, and adding a corresponding ``hole" part.

In fact, we can sketch a simple model for the condensed matter system above 
that generates the same qualitative picture as the abelianization of the ABJM model.
Consider spinless bosons $\phi_{ij}$ generated by coupling fermions of opposite spins (Cooper pairs) at sites $i$ and $j$
with a maximum distance between sites $|i-j|\leq N$, {\i.e.} $\bar \psi_i \psi_j$. The resulting $\phi_{ij}$ can be described by a field 
$\phi_{i'}^{ab}$, with $a,b=1,...N$. Since we are in two spatial dimensions, every site has 
${\cal O}(N^2)$ neighbours a distance $\leq N$ away. 
Now take the point $i'$ at which the effective field, $\phi_{ij}$, lives to be midpoint of the line between $i$ and $j$, and $a,b$ to correspond to sites $j$ in the $x$ and $y$ directions away from $i'$ (so that, if $i'$ and $j$ are fixed, so is $i$). Consider that 
the normalized wavefunctions for the field $\phi_{i'}^{ab}$ give probabilities for existence of the pairing as $\propto |\phi_{i'}^{ab}|^2$ for a 
pair $(ab)$. In this case, any transformation on $\phi_{i'}^{ab}$ must be a unitary transformation $U^{ab,a'b'}$ inside $U(N^2)$, up to an overall factor. In particular, any {\it symmetry} 
of the system must be of this type. The symmetry of the ABJM model is $U(N)\times U(N)$, and would correspond to $U^{ab,a'b'}=f U^{aa'}V^{bb'}$.

Since the simplest type of condensed matter system is a rotationally invariant one, we should not have any angular dependence, and 
we should have $\phi^{ab}_i=\phi_i(\sqrt{a^2+b^2})$ $=\phi_i^{ba}$. It should be then 
possible to diagonalize this symmetric matrix, corresponding to 
considering only the constant (rotationally invariant) $m=0$ modes for the "spherical harmonics" expansion $e^{2\pi i m\theta}$ at fixed radius 
$r=\sqrt{a^2+b^2}$. 
In this way, only ${\cal O}(N)$ modes, specifically those that are spherically symmetric, out of the ${\cal O}(N^2)$ modes in the system are turned on. These can be thought of as the eigenvalues of $\phi_i^{ab}$.

Since $N$ is the effective maximal radius for coupling of the two fermions at different sites, it makes sense 
for the wavefunction in the ground state to decrease from a 
maximum value at $a=1$ (neighbouring sites) to zero at $a=N$ (sites at distance $N$). For instance, if the wavefunction $\psi(a)$ is such that 
$|\psi(a)|^2\propto N-a$, then the average distance between sites is 
\be
  \langle a  \rangle=\frac{\int |\psi(a)|^2 a (2\pi a da)}{\int |\psi(a)|^2(2\pi a da)}=\frac{N}{2}\,,
\ee
which is consistent with having a large average distance between the electrons that couple. 
This form of the wavefunction, $\psi(a)\propto \sqrt{N-a}$, here just a consistent choice, is exactly what we obtain in the ABJM model. Of course, in principle, if we would be able to correctly describe the interactions between various $\psi_{i'}^{ab}$, as in the ABJM model, 
the dynamics would select the form of $\psi(a)$. Finally, the Hubbard model field $b_i$ must be 
the linear combination of the spherical modes, i.e. $b_i\sim \sum_a\psi(a)\phi_i^{aa}$.
 
We have already seen that to obtain the Landau-Ginzburg model from ABJM, we have only one field, $\chi_2$, turned on corresponding to turning on the matrix $G^2=\sqrt{N-m}\delta_{m+1,n}$, with $(G^2G^\dagger_2)_{mn}=(N-m)\delta_{mn}$. As in the simple model above, there are two independent rotations, in this case $U(N)\times U(N)$
rotations, acting on the indices, so the most general solution for the matrix $G^2$ is in fact $U(\sqrt{N-m}\delta
_{mn})V^{-1}$. We can use these to diagonalize the matrix, thus reducing 
the degrees of freedom turned on, from ${\cal O}(N^2)$ to ${\cal O}(N)$, as in the above condensed matter model.
The ABJM field that is turned on is $\chi_2(G^2)_{mn}$, corresponding to $b_{i'}\sim\sum_a \psi_a \phi_{i'}^{aa}$.

While this field is the only one turned on in our simple toy condensed matter model,  there are, in principle, many 
more fields. We could, for instance, have more free electrons at each site, thus having more matrix 
scalars, transforming in some R-symmetry group (in ABJM we have 4 complex scalars, corresponding 
to $\phi_1,\phi_2,\chi_1,\chi_2$, that transform under the $SU(2)\times SU(2)$ of the mass deformed ABJM). Then , we could also have matrix fermions, corresponding for instance to two electrons at site $i$ coupling with one electron at site $j$, although such modes are, of course, not turned on in the Hubbard model description. To complete the field content of the \
ABJM model we need also the Chern-Simons gauge fields, but since those are topological and have no dynamics, 
we don't need to introduce any new degrees of freedom.

Chern-Simons gauge fields are, of course, no strangers to condensed matter systems, showing up, for instance, in the fractional quantum Hall effect (see for instance the review \cite{simon}). An abelian Chern-Simons field can be obtained as follows. Consider a multi-electron wavefunction $\Psi_e(\vec{r}_1,...,\vec{r}_k)$ with a generic Hamiltonian
\be
  H_e=\sum_j\frac{|\vec{p}_j-e\vec{A}(\vec{r}_j)|^2}{2m_b}+\sum_{i<j}v(\vec{r}_i-\vec{r}_j)
\ee
such that $H_e\Psi_e=E\Psi_e$. We can redefine the wavefunction through the transformation
\be
  \Phi(\vec{r}_1,...,\vec{r}_k)=U\Psi_e(\vec{r}_1,...,\vec{r}_k)=
  \left[\prod_{i<j}e^{-i\tilde \phi\a(\vec{r}_i-\vec{r}_j)}\right]\Psi_e(\vec{r}_1,...,\vec{r}_k)
\ee
where $\a(\vec{r}_i-\vec{r}_j)$ is the angle made by $\vec{r}_{ij}=\vec{r}_i-\vec{r}_j$ with a fixed axis. Since 
\be
  U^{-1}(\vec{p}_i-e\vec{A}(\vec{r}_i))U=\vec{p}_i-e\vec{A}(\vec{r}_i)-e\vec{a}(\vec{r}),
\ee
where
\be
  e\vec{a}(\vec{r}_i)=\tilde\phi\sum_{j\neq i}\vec{\nabla}_i\a(\vec{r}_i-\vec{r}_j),\label{csgf}
\ee
the new Hamiltonian reads
\be
H=\sum_j\frac{|\vec{p}_j-e\vec{A}(\vec{r}_j)-e\vec{a}(\vec{r}_i)|^2}{2m_b}+\sum_{i<j}v(\vec{r}_i-\vec{r}_j)
\ee
so that $H\Phi=E\Phi$. Therefore after the transformation, $\vec{a}(\vec{r})$ describes a gauge field with no dynamics which, one can show is of Chern-Simons type. Such a Chern-Simons gauge field, coupled to the fermions and to the electromagnetic gauge field, plays a central role in the fractional quantum Hall effect, see e.g. \cite{witten}.

A generalization of this construction to the nonabelian case is straightforward. If two fermions at sites $i$ and $i''$ couple to form a boson
$\phi_{i'}^{aa'}$, at site $i'$ at the midpoint, and two other fermions at sites $j$ and $j''$ couple to form a boson $\phi_{j'}^{bb'}$ at 
site $j'$ at their midpoint, we can consider the field
\be
  e\vec{a}\left(\vec{r}_{i'}\right)=\vec{\nabla}_{i'}\sum_{j'\neq i'}\a\left(\vec{r}_i-\vec{r}_j\right)\,,
  \label{nonabcs}
\ee
where we have not yet specified the nonabelian indices on the gauge field. It is not hard to see that the only variable in this object is 
the vector $\vec{r}_{ii'}-\vec{r}_{jj'}$ (by changing the vector $\vec{r}_{ii'}$ we just produce a harmless global spatial translation in the value 
of the right hand side of (\ref{nonabcs})), as well as the discrete 
choice of $\vec{r}_{ii'}$ to belong to the fixed point $i'$ or the summed point $j'$.
Since the two vectors $\vec{r}$ subtracted correspond to matrix indices $(aa')$ and $(bb')$, 
we can think of this construction as giving us two nonabelian gauge fields 
$\vec{a}^{ab}$ and $\hat{\vec{a}}^{a'b'}$, like the $A$ and $\hat A$ of ABJM. Moreover, the scalars $\phi_{i'}^{aa'}$ are bifundamental with 
respect to the two resulting gauge fields. There remain many open problems to understand about this model, not the least of which is the symmetry group acting on the matrix Chern-Simons fields but we leave these to the interested reader. This concludes our description of the field content of ABJM and qualitative undestanding of abelianization.
Suffice it to say that the ABJM abelianization gives a well motivated model of AdS/CMT.

Finally, a few comments on a four dimensional picture for the Landau-Ginzburg model (\ref{lgpot}). The Landau-Ginzburg model makes more sense 
from a theoretical viewpoint as a 
reduction of the corresponding four dimensional theory. But here as well, the abelianization presented has in particular an ansatz with the 
scalar VEV $b$ multiplying the matrix $G^1$. If we had the same VEV multiplying both $G^1$ and $G^2$, that would lead to a description of the fuzzy two-sphere, a finite $N$ approximation of the clasical two-sphere \cite{Nastase:2009ny,Nastase:2010uy}. As it is, we can think of the abelianization 
as generating a single direction, or a "fuzzy circle", therefore the resulting Landau-Ginzburg theory must also be thought of as coming from a circle reduction of a similar theory in 4 dimensions. The physical radius obtained from a fuzzy space construction was argued to be (see for instance \cite{Nastase:2009ny})
\be
  R_{ph}^2=\frac{2}{N}\Tr\left[X^IX_I^\dagger\right]=\frac{2}{N}\Tr\left[C^IC^\dagger_I\right]4\pi^2l_P^3\,,
\ee
where $l_P^3=l_s^2R_{11}$. Assuming this same formula holds for the less defined ``fuzzy circle" case, from 
$\Tr[G^1G^\dagger_1]=N(N-1)/2$, we get
\be
  R_{ph}^2=(N-1)|b|^24\pi^2l_s^2R_{11}\,.
\ee
If, as in the pure fuzzy sphere case, the 11th direction has radius $R_{11}=R_{ph}/k$, we obtain in the ``maximally Higgs" case $|b|^2=c^2$
\be
  R_{ph}=(N-1)\mu l_s^2\,.
\ee

\section{Some BPS solutions with spacetime interpretation}

We now return to the more general abelianization ansatz, and
consider the system with $\phi_1=\phi_2=0$ and  gauge fields put to zero, but $\chi_1\neq 0$ still a field (unlike in the abelian Higgs case previously described). This ansatz gives the reduced action
\bea
  S&=&-\frac{N(N-1)}{2}\int d^{3}x\left[|\d_{\mu}\chi_{1}|^{2}+|\d_{\mu}\chi_{2}|^{2}\right.\cr
  &&\left.+\frac{4\pi^{2}}{k^{2}}\Big(|\chi_{1}|^{2}\big(|\chi_{2}|^{2}-c^{2}\big)^{2}
  +|\chi_{2}|^{2}\big(|\chi_{1}|^{2}-c^{2}\big)^{2}\Big)\right]
  \label{redact}
\eea
which we now proceed to study. 

\subsection{Single Profile Solution}
As a first pass, let's consider solutions with a single profile
\be
  \chi_1=\chi_2=f(x_1)\,,
\ee
with $x_1$ as one of the spatial coordinates. The equation of motion for the reduced action (\ref{redact}) for this ansatz become
\be
  \partial_{x_{1}}^{2}f=\frac{4\pi^{2}}{k^{2}}f\left(f^{2}-\frac{\mu k}{2\pi}\right)\left(3f^{2}-
  \frac{\mu k}{2\pi}\right)\,,
  \label{domainwalleom}
\ee
from which we distill two cases:
\begin{enumerate}
  \item \underline{Zero mass:} In the massless case, $\mu=0$, the ground state solution is simply $f(x_1)=0$ 
  with no other constant solutions.  This is, however, not the only solution and a straightforward integration of 
  the equation of motion yeilds 
  \be
    f(x_{1})=\sqrt{\frac{k}{4\pi x_{1}}}\,.\label{fzfnl}
  \ee
  This is a {\it fuzzy funnel solution} which we can check is, in fact, BPS. Indeed, the energy of 
  solutions satisfying the above ansatz is
  \be
    H_{\mu=0}=-N(N-1)\int dx_{1}dx_{2}\left[(\partial_{x_{1}}f)^{2}+\frac{4\pi^{2}}{k^{2}}f^{6}\right]\,,
  \ee
  which, by the usual procedure of completion of squares, can be expressed as 
  \be
    H_{\mu=0}=-N(N-1)\int dx_{1}dx_{2}\left(\left(\partial_{x_{1}}f-\frac{2\pi}{k}f^{3}\right)^{2}+\rm surface\;    
    term\right)\,,
  \ee
  from which we can simply read off the BPS equation
  \be
   \partial_{x_{1}}f(x_{1})=\frac{2\pi}{k}f(x_{1})^{3}\,.
  \ee
  It is clear that this equation is solved by the fuzzy funnel solution above.

  \item \underline{Nonzero mass:} In this case, the constant solutions to the equations of motion are
  \be
    f=0,\;\;\;
    f=\sqrt{\frac{\mu k}{2\pi}},\;\;\;
    f=\sqrt{\frac{\mu k}{6\pi}}\,.
  \ee
  Of these, only the first two are ground states. Indeed, completing squares again, we find the BPS equation
  \be
   \partial_{x_{1}}f+\frac{2\pi}{k}f\left(f^{2}-\frac{\mu k}{2\pi}\right)=0\,,
  \ee
  from which see that indeed $f=0$ is a trivial ground state, while the second solution ($f^2=\mu k/2\pi$) is 
  again the fuzzy sphere ground state. The third solution of the equations of motion ($f^2=\mu k/6\pi$) doesn't    
  satisfy the BPS equation, so is a non-ground state fuzzy sphere. The BPS equation has nontrivial solutions 
  \bea
   f_{\mp}(x_{1})&=&\sqrt{\frac{\mu k/2\pi}{1\mp e^{-2\mu x_{1}}}}\,.
  \eea
  The first solution, $f_{-}$, describes a fuzzy funnel with $x_1\in(0,+\infty)$, so $f_{- }$ varies between an   
  infinite size at $x_{1}=0$ and the fuzzy sphere ground state at $x_{1}\to+\infty$,
  \be
    f_{-}(0)=+\infty,\;\;\;\;
    f_{-}(+\infty)=\sqrt{\frac{\mu k}{2\pi}}\,.
  \ee
  The second solution, $f_{+}$, describes a fuzzy funnel with $x_1\in (-\infty,+\infty)$, varying in size between 
  zero at $x_{1}\to-\infty$ and the fuzzy sphere at $x_{1}\to+\infty$, 
  \be
    f_+(-\infty)=0, \;\;\;\;
    f_+(+\infty)=\sqrt{\frac{\mu k}{2\pi}}\,.
    \label{f2}
  \ee
  This fuzzy funnel solution will be elaborated on in the next section, where we argue that it is a 
  generalization of the Basu-Harvey solution that describes an M2 ending on a {\it spherical} M5. These solutions
  are plotted in figure 4. below.
  \begin{figure}[h]
    \centering
    \includegraphics[width=5cm,height=4cm]{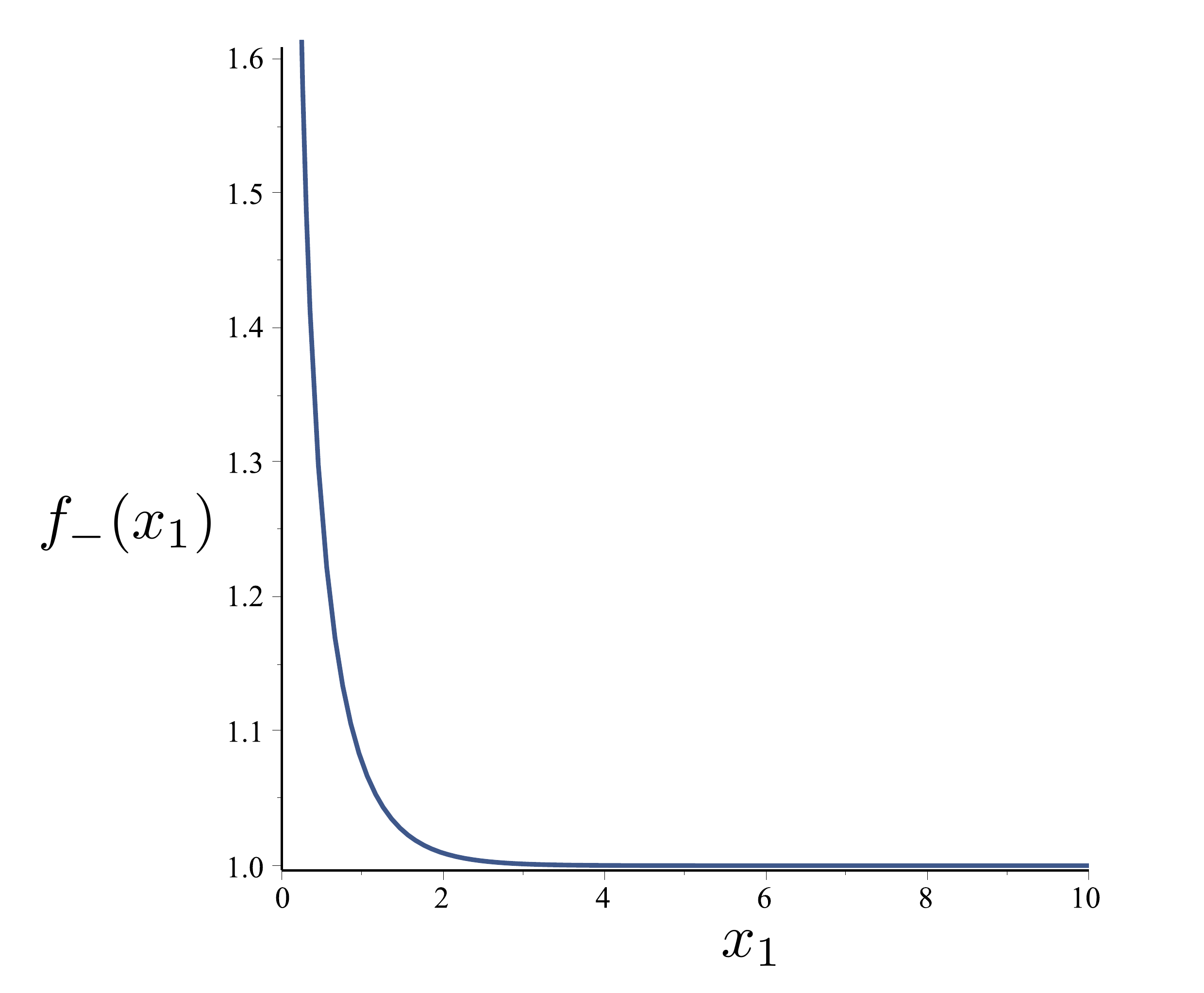}
    \includegraphics[width=5cm,height=4cm]{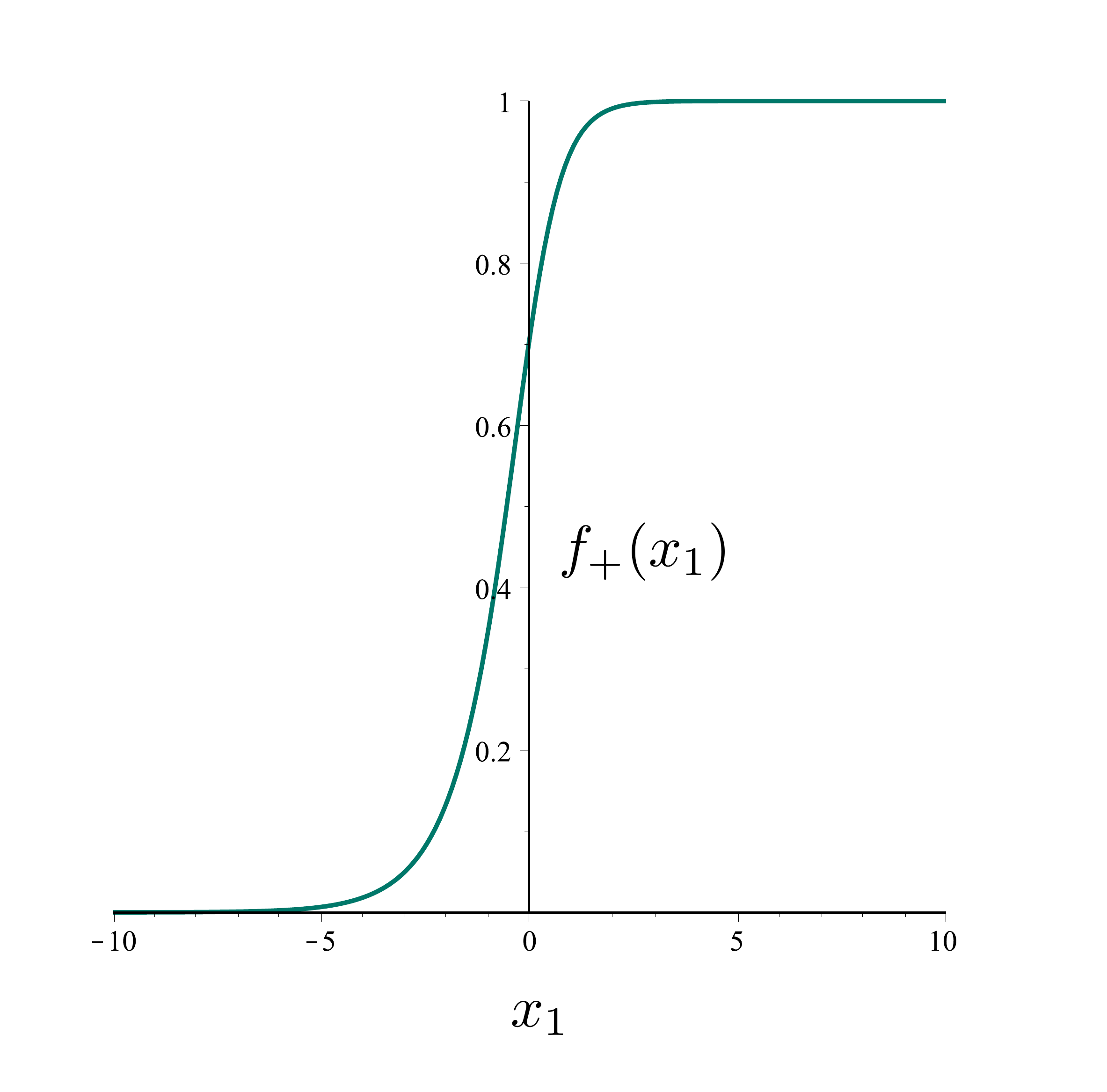}
    \caption{The normalized single-profile solutions $f_{-}(x_{1})$ and $f_{+}(x_{1})$}
  \end{figure}  
\end{enumerate} 
\subsection{Two-Profile Solution}
The above single profile solution is also fairly easily generalized to a two-profile one with
\be
  \chi_1=f(x_1),\;\;\;\;
  \chi_2=g(x_1)\,.
\ee
With this ansatz, the equations of motion reduce to 
\bea
  \partial_{x_{1}}^{2}f&=&\frac{4\pi^{2}}{k^{2}}f\Big[\Big(g^{2}-\frac{\mu k}{2\pi}\Big)^{2}+2g^{2}\Big(f^{2}-  
  \frac{\mu k}{2\pi}\Big)\Big]\,,\nonumber\\
  \\
  \partial_{x_{1}}^{2}g&=&\frac{4\pi^{2}}{k^{2}}g\Big[\Big(f^{2}-\frac{\mu k}{2\pi}\Big)^{2}+2f^{2}\Big(g^{2}-  
  \frac{\mu k}{2\pi}\Big)\Big]\,.\nonumber
\eea
Again, we can complete squares in the Hamiltonian and read off the BPS equations
\bea
  \partial_{x_{1}}f+\frac{2\pi}{k}f\Big(g^{2}-\frac{\mu k}{2\pi}\Big)&=&0\,,\nonumber\\
  \\
  \partial_{x_{1}}g+\frac{2\pi}{k}g\Big(f^{2}-\frac{\mu k}{2\pi}\Big)&=&0\,.\nonumber
  \label{BPSfg}
\eea
As in the single profile case above, there are again two separate cases that need to be solved separately.
\begin{enumerate}
  \item \underline{Massless case:} For $\mu=0$, we have the solutions
  \bea
    f(x_{1})&=&\sqrt{\frac{k}{2\pi}}\frac{Ce^{-x_{1}}}{\sqrt{1-C^2e^{-2x_{1}}}}\,,\nonumber\\
    \\
    g(x_{1})&=&\sqrt{\frac{k}{2\pi}}\frac{1}{\sqrt{1-C^2e^{-2x_{1}}}}\,,\nonumber
  \eea
   that solve both the first order BPS equations of motion as well as the general second order equations. 
   This solution blows up at $x=\log C$, and goes to a constant in $g$ and zero in $f$, 
   corresponding to a fuzzy circle. These solutions are plotted in figure 5. below.
   \begin{figure}[h]
  \centering
  \includegraphics[width=5cm,height=4cm]{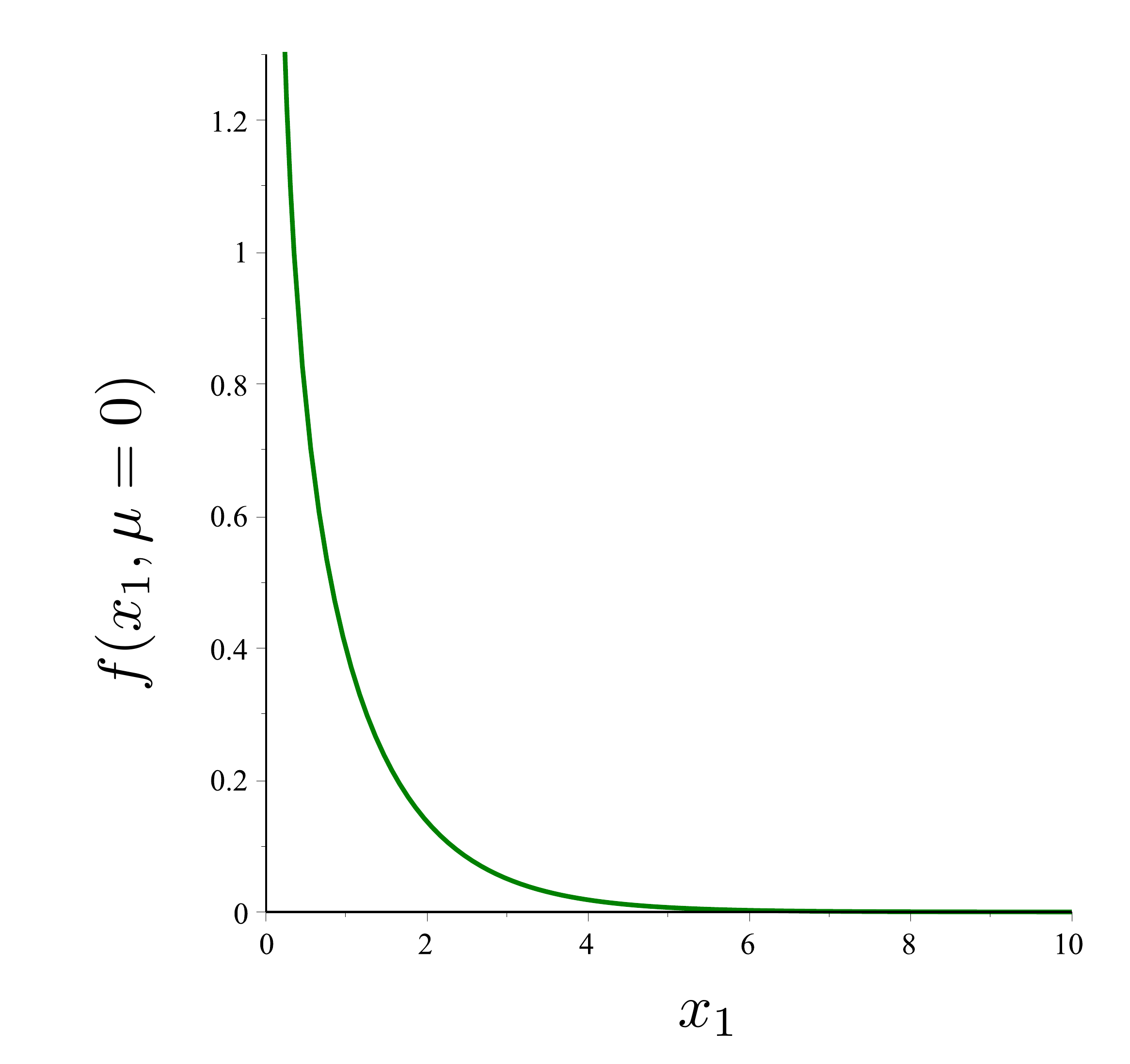}
  \includegraphics[width=5cm,height=4cm]{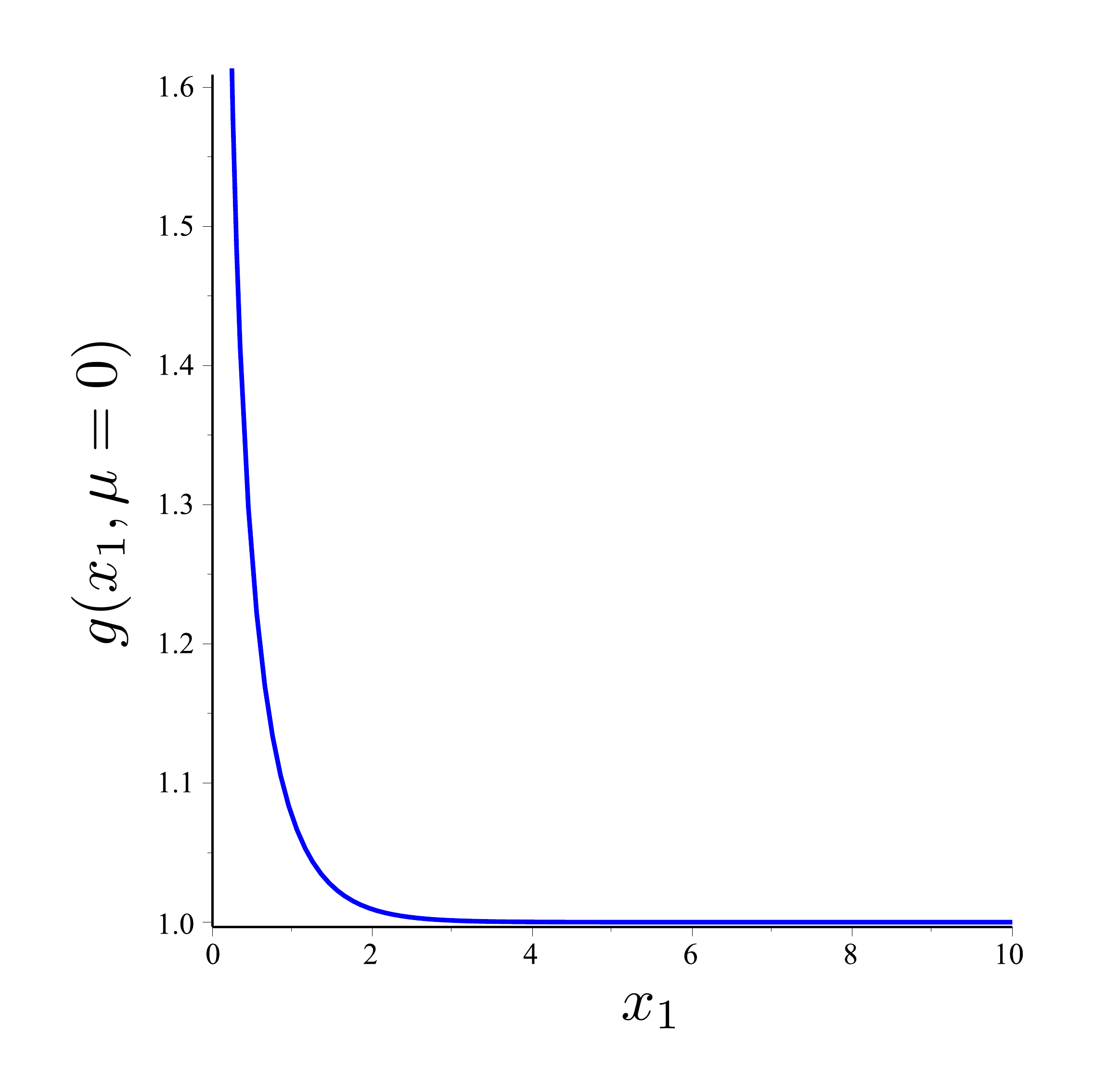}
  \caption{The normalized two-profile solutions $f(x_{1})$ and $f(x_{1})$}
\end{figure}  

  \item \underline{Nonzero mass:} For $\mu\neq 0$, we have the solutions (see figure 6)
  \bea
    f(x_{1})&=&\sqrt{\frac{\mu k}{2\pi}}
    \frac{C\exp\big[\mu x_{1}-\frac{1}{2}\rm e^{2\mu x_{1}}\big]}{\sqrt{1-   
    C^2\exp(-\rm e^{2\mu x_{1}})}}\,,\nonumber\\
    \\
    g(x_{1})&=&\sqrt{\frac{\mu k}{2\pi}}\frac{\rm e^{\mu x_{1}}}{\sqrt{1-C^2\exp(-\rm e^{2\mu x_{1}})}}
    \,.\nonumber
  \eea
  \begin{figure}[h]
  \centering
  \includegraphics[width=5cm,height=4cm]{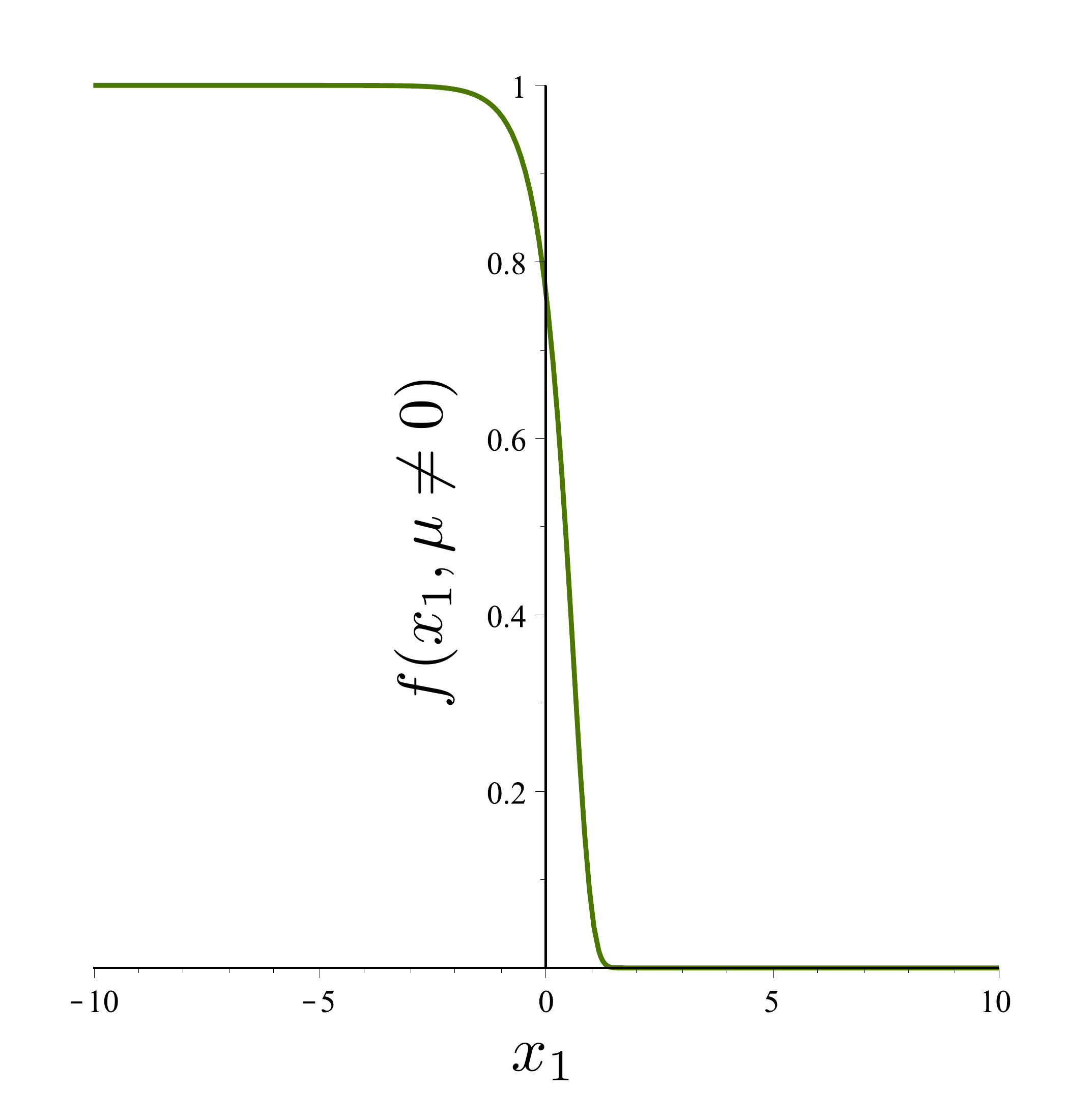}
  \includegraphics[width=5cm,height=4cm]{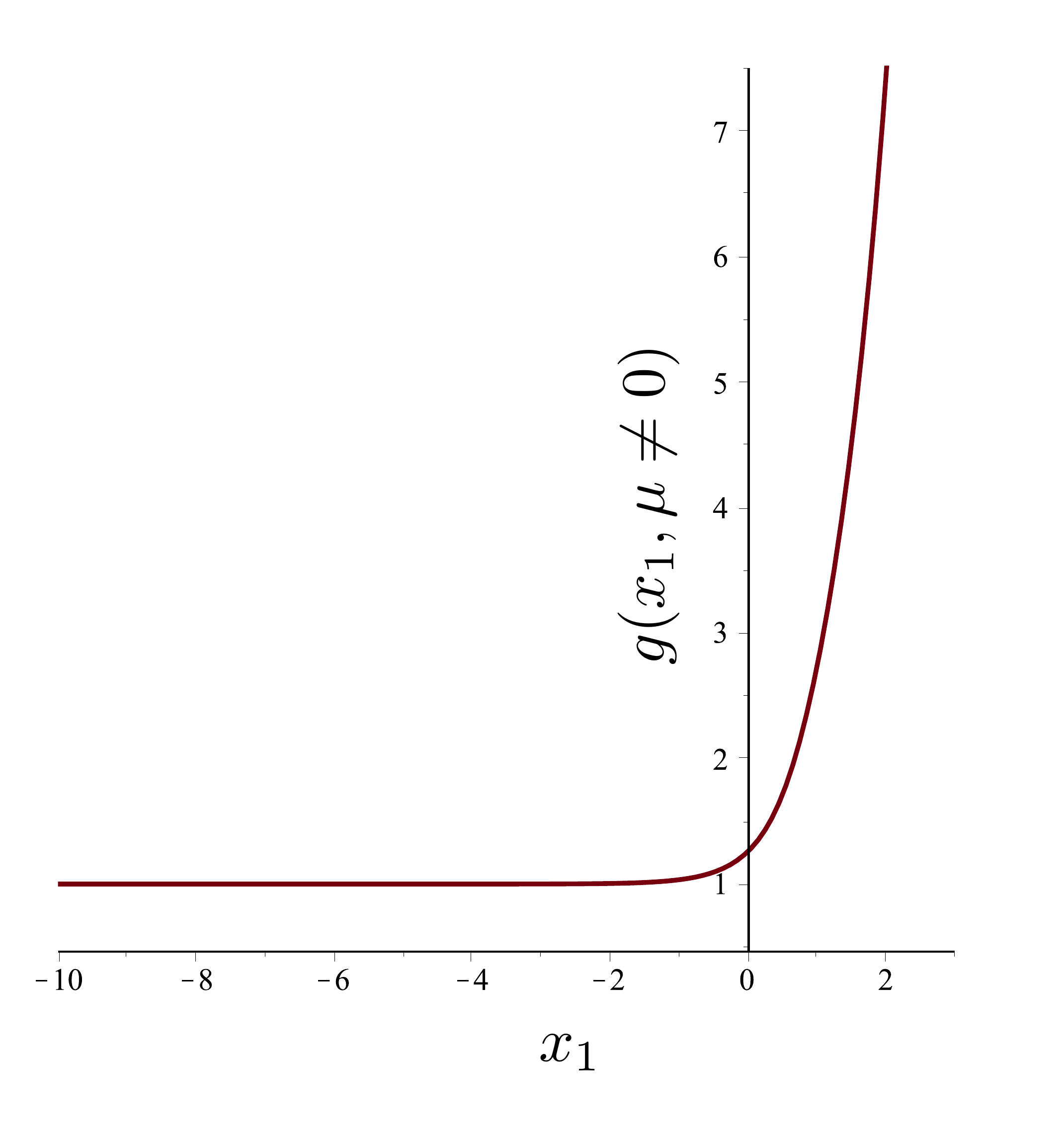}
  \caption{The normalized two-profile solutions $f(x_{1})$ and $g(x_{1})$ for $\mu\neq 0$.}
\end{figure}  

\end{enumerate}

\section{Funnel solutions as M2-M5 brane systems}

In this section we will try to find a spacetime interpretation for the fuzzy funnel solutions in eq. (\ref{f2}). 
The fuzzy funnel solution (\ref{fzfnl}), interpolating between a sphere of infinite size and a sphere of zero size, 
is known to have the spacetime interpretation of a flat M2-brane ending on a flat M5-brane. From the point 
of view of the M2-brane theory given by the massless ABJM, the M5-brane appears as a spherical funnel solution, a M5-brane that grows from zero size at 
$x_1=\infty$ to infinite size at $x_1=0$. We will review this case, reduced to string theory, i.e. a D2-brane ending on a D4-brane, later.
Also, from the point of view of the M5-brane theory, we can write a BIon type solution, corresponding to an M2-brane growing out of the M5-brane
(directions 0 and 5 are trivial, and in directions 1-4 the M2-brane appears as a BIon). From the point of view of the spacetime
theory, we have an M2-M5 system preserving 1/4 supersymmetry in flat space, and we also have an M5-brane solution in the (backreacted) background of 
M2-branes. This picture matches nicely with the two worldvolume descriptions.

With this in mind, we expect that the fuzzy funnel solution of (\ref{f2}) should have a similar interpretation. The solution interpolating between zero and a fuzzy sphere vacuum was found in \cite{Hanaki:2008cu,Arai:2008kv}, and we would guess that it can only match 
with a spacetime solution corresponding to an M2-brane ending on 
an M5-brane. We will see however that there is some ambiguity, related to the existence of two solutions, the one from zero to the fuzzy sphere, and one from the fuzzy sphere to infinity.

\subsection{Massless case: A Fuzzy Funnel Review}

The solution (\ref{fzfnl}) corresponds in spacetime to a flat M2-brane ending on a flat M5-brane, a solution which preserves 1/4 supersymmetry 
as follows. In a flat background, the 1eleven dimensional gravitino transformation law, 
\be
  \delta \psi_\mu=D_\mu\epsilon+\# ({\Gamma^{\nu\rho\sigma\lambda}}_\mu-8\delta_\mu^\nu \Gamma^{\rho  
  \sigma\lambda})F_{\nu\rho\sigma\lambda}\,,
\ee
must be set to zero in order to obtain a BPS solution. The M2-brane solution extended in the (0,1,2)-directions 
corresponds to a nonzero 3-form $A_{012}$, with a nonzero field strength component $F_{012r}$ 
(here $r$ is the radial part of all the coordinates transverse to the M2). 
The solution is given by a local supersymmetry parameter $\epsilon(r)$ which is a scalar function of $r$ times a constant susy parameter
$\epsilon_{(0)}$ satisfying 
\be
  \Gamma^{012}\epsilon_{(0)}=\pm \epsilon_{(0)}\,.
\ee
The M5-brane solution extended in the $(0,1,3,4,5,6)$ directions similarly gives a nonzero field strength $F_{\theta_1...\theta_4}$, where $\theta_1,...,\theta_4$ are the four angles obtained for the transverse directions $(2,7,8,9,10)$. Again, the solution for the local supersymmetry parameter $\epsilon(r)$ is a function of $r$ times a constant susy parameter $\epsilon_{(0)}$ satisfying
\be
  \Gamma^{013456}\epsilon_{(0)}=\pm \epsilon_{(0)}\,.
\ee
We can then have a solution for an M2-brane ending on an M5-brane preserving 1/4 supersymmetry by imposing both conditions (which are now compatible). We can then reduce this system to 10-dimensional string theory, thereby considering a D2-brane ending on a D4-brane.

From the point of view of the D4-brane theory in a flat background spacetime, the fuzzy funnel solution looks like a BIon-type solution. For a spacetime D2-brane in the (0,1,2)-directions, called $t,x,z$, and a D4-brane in the (0,1,3,4,5)-directions, with polar coordinates $r,\theta,\phi$ for the directions (3,4,5), the worldvolume gauge field flux on the D4-brane is 
\be
  F=(2\pi\a')n \sin\theta d\theta d\phi\,.
\ee
Because the solution is of the BIon type, with the D2-brane growing out of the D4-brane, we consider $z=z(r)$ on the worldvolume, leading to DBI D4-brane Lagrangian 
\be
  {\cal L}=T_4\sqrt{(1+z'(r)^2)(r^4+(\lambda n)^2)}\,,
\ee
where $\lambda\equiv 2\pi \a'$. Since ${\cal L}$ is independent of $z$, it follows that $\partial {\cal L}/\partial z'$ is a constant, 
which we can  put equal to $\lambda n$, in which case we obtain 
\be
z'=\pm \frac{\lambda n}{r^2}\Rightarrow z=\frac{\lambda n}{r}
\ee
This corresponds to a funnel solution for a semi-infinite D2-brane ending on the D4-brane.

A similar story takes place for the case where there is a background created by other D2-branes (parallel with the first). It is however easier to describe what happens in the case of the type IIB solution for D3-branes ending on D5-branes (instead of D2-D4), 
since in that case the spacetime background is easier (there is no M theory reduction). 
Consider the background generated by other D3-branes, with harmonic function $f(r)$,
\bea
  ds^2&=&f(r)^{-1/2}(-dt^2+dx^2+dy^2+dz^2)+f(r)^{+1/2}(dr^2+r^2d\Omega_2^2+d\vec{s}^2)\nonumber\\
  \\
  C_{(4)}&=&(f^{-1}-1)dt\wedge dx\wedge dy\wedge dz\nonumber
\eea
The DBI Lagrangean for the D5-brane reads
\be
  {\cal L}=T_5\left[\sqrt{(1+f'(r)z'(r)^2)(r^4+f^{-1}(r)\lambda^2n^2)}-\lambda n(f(r)^{-1}-1)z'(r)\right]\,,
\ee
and the same calculation leads to the same solution $z(r)=\lambda n/r$, with the function $f(r)$ dropping out completely. One can also take the 
near-horizon limit and consider the usual scaling $r=\a' U$, leading to a finite funnel solution that can be interpreted from the point of view of the D3-brane theory as
\be
  U(z)=\frac{n}{2\pi z}\,.
\ee

Note that if we consider a spherical D5-brane ansatz, oriented in the $(t,x,y,z,\Omega_2)$-directions, we obtain the Lagrangian
\be
  {\cal L}=T_5\sqrt{f^{-1}(r)(r^4+g^{-1}(\lambda n)^2)}-\lambda n \left(f^{-1}(r )-1\right)\,.
\ee
If we take the full harmonic function $f( r)=1+\frac{Q}{r^4}$,
there is no fixed sphere solution with $r=R=$constant, 
but if we drop the 1 in $f$, i.e. at very large $r$, we obtain an identity by varying with respect to 
$r$, namely $\lambda n/Q-\lambda n/Q=0$. Therefore in flat space, we have asymptotically a solution for very large radius sphere, but at small radius we only have the funnel solution; a fixed sphere is not a solution.

\subsection{Massive case: Supersymmetry and a Fluctuation Solution on the M5-brane}

The mass deformation changes the 11 dimensional background spacetime from flat to \cite{Mohammed:2010eb,Lambert:2011eg}
\bea
  ds^2&=&H^{-2/3}(-dt^2+dx_1^2+dx_2^2)+H^{1/3}(dx_3^2+...dx_{10}^2)\nonumber\\
  \\
  F_4&=&2\mu (dx^3\wedge dx^4\wedge dx^5\wedge dx^6+dx^7\wedge dx^8\wedge dx^9\wedge dx^{10})+dx^0\wedge   
  dx^1\wedge dx^2\wedge dH^{-1}\nonumber
  \label{massdef}
\eea
where $H(r)=1-\frac{1}{4}\mu^2r^2$. A naive guess is that the M5-brane has to live in the $(0,1,2,\theta,\phi,\xi)$- directions, where $\theta,\phi$ and$\xi$ are the angular 
directions of $(3,4,5,6)$, with $r$ their radial direction, giving
\be
  \Gamma^{012\theta\phi\xi}\epsilon_{(0)}=\pm \epsilon_{(0)}\,,
\ee
so that the transverse M2-brane would have to be in the $(0,1,r)$-directions, giving
\be
  \Gamma^{01r}\epsilon_{(0)}=\pm \epsilon_{(0)}\,.
\ee
However we observe that the 4-form in (\ref{massdef}) has nonzero $F_{012r}$ as wanted, but since $F_{012r}=\d_r(A_{012})$ and not $\d_2(A_{01r})$, there must be a nontrivial Maxwell transformation that brings the gauge field $A$ into the desired form.

How would this M2-M5-brane solution look from the point of view of the D4-brane theory (i.e., reducing to 10d string theory and focusing on the worlvolume theory)? 

From the fuzzy $S^2$ picture in the ABJM theory an action was found for the fluctuation modes around the ground state 
\cite{Nastase:2009ny}. For the scalar $\Phi$ 
corresponding to the fluctuations of the radius of the $S^2$ (transverse direction) the action reduces to just a massive mode, i.e. with 
potential
\be
V_\Phi=\frac{1}{2}\left[\Phi^2+(\nabla_{S^2}\Phi)^2\right]\,,
\ee
Later, in \cite{Lambert:2011eg} 
the same fluctuation action was found from the DBI action of a D4-brane in the background (\ref{massdef}).

For such a potential, a solution was found in \cite{Sadri:2003mx}, 
and was called the BIGGon (in analogy with the BIon), 
representing, in spacetime, an $S^3$ giant graviton in type IIB on the maximally supersymmetric pp wave background with F-string spikes attached at the poles. 
Since the pp-wave background is T-dual to (\ref{massdef}) (see, for example, \cite{Mohammed:2010eb} for the explicit construction), the same solution should apply in our case. The BIGGon was found by similarly taking a single scalar field $\Phi$ on the 3-sphere (a fluctuation of the radial coordinate), with the same potential, but on $S^3$ instead of $S^2$, {\it i.e.}
\be
  V_\Phi=\frac{1}{2}\left[\Phi^2+(\nabla_{S^3}\Phi)^2\right]\,.
\ee
The solution is
\be
\Phi=\frac{Q}{\sin\psi}\,,
\ee
giving the full radial coordinate (background plus fluctuation) 
\be
X=R\left(1+Q\frac{g_{eff}}{\sin\psi}\right)\,,
\ee
where we have taken the $S^3$ parametrization to be 
\bea
X^4&=&R\cos\psi\,,\nonumber\\
X^3&=&R\sin\psi\cos\theta\,,\nonumber\\
X^2&=&R\sin\psi\sin\theta\sin\phi\,,\nonumber\\
X^1&=&R\sin\psi\sin\theta\cos\phi\,.\nonumber
\eea
Therefore in our case we have the BIGGon solution 
\be
\Phi=\frac{Q}{\sin\theta}
\ee
where the parametrization of the $S^2$ is
\bea
X^3&=&R\sin\theta\,,\nonumber\\
X^4&=&R\cos\theta\sin\phi\,,\nonumber\\
X^5&=&R\cos\theta\cos\phi\,.\nonumber
\eea
This solution indeed corresponds with our naive expectation of a D2-brane extending out perpendicularly from the spherical D4-brane. But the action that it extremizes corresponds to small fluctuations of the field $\Phi$. However, in \cite{Lambert:2011eg} it was shown how to write the full DBI action for D4-branes in the mass-deformed spacetime.

{\bf Massive case: full funnel solution}

In the background (\ref{massdef}) it was shown that the DBI action for the D4-brane has a fixed sphere solution, corresponding to the fuzzy sphere 
solution of the massive ABJM. Now we want to see if we can also find funnel solutions corresponding to the ABJM solutions (\ref{f2}) and extending
the perturbative BIGGon solution above. 

To this end, we consider again an M2-brane in the (0,1,2)-directions and an M5-brane in the (0,1,3,4,5,6)-directions, with (3,4,5,6) in polar coordinates $r,\theta,\phi$, and $\xi$. We reduce M-theory to type IIA on $\xi$ and look for a D4-brane extending along $0,1,r,\theta,\phi$, with $z=z(r,\theta,\phi)$ in order to have 
a BIon-type solution corresponding to a perpendicular D2-brane as above.

Dimensionally reducing the background (\ref{massdef}) to type IIA string theory, and writing only the terms in directions parallel to the D4-brane,
we find \cite{Lambert:2011eg}
\bea
  C_{(5)}&=&-\frac{\mu}{2kR_*}\left(\frac{H^{-1}+1}{2}\right)dt\wedge dx\wedge z'dr\wedge r^4d\Omega_2+...
  \nonumber\\
  C_{(3)}&=&(H^{-1}-1)dt\wedge dx\wedge z' dr+É\\
  B&=&\frac{\mu}{2kR_*}r^4d\Omega_2+É\nonumber\\
  e^\phi&=&\left(\frac{r}{kR_*}\right)^{3/2}H^{1/4}\nonumber
\eea
and a worldvolume flux $F=2\lambda N d\Omega_2$, giving
\be
  {\cal F}=\lambda F-B=2\lambda N-\frac{\mu }{2kR_*}r^4\,.
\ee
Substituting this ansatz into the action
\be
  S=T_4\left[\int d^5x e^{-\phi}\sqrt{-\det(g+{\cal F})}+\int (C^{(5)}+C^{(3)}\wedge {\cal F})\right]\,,
\ee
gives
\bea
  S&=&T_4\left\{\int dt\wedge dx\wedge dr\wedge d\Omega_2\sqrt{(1+z'^2H^{-1})\left(1+H^{-1}\frac{k^2R_*^2}  
  {r^6}\left(2\lambda N
  -\frac{\mu r^4}{2kR_*}\right)^2\right)}\right.\cr
  &+&\left.\int dt\wedge dx\wedge dr\wedge d\Omega_2\left(2\lambda N z'(H^{-1}-1)-\frac{\mu }{2kR_*}  
  z'\frac{H^{-1}+1}{2}r^4\right)\right\}\,.
\eea
Evidently, the Lagrangian ${\cal L}$ is independent of $z$, which means that $\partial L/\partial z'$ is a constant, which we can put equal to $-T_42\lambda N$, in which case  
\be
 z'=\pm\frac{\Big[\frac{2\lambda N}{r^3}-\frac{\mu r}{2}+\frac{\mu^3r^3}{16}\Big]\sqrt{1-\frac{\mu^2r^2}  
 {4}}}{\sqrt{1-\frac{\mu^2r^2}{4}-\frac{\mu^6
  r^6}{4^4}+\frac{\mu^4r^4}{4^2}-\frac{\mu^3kR_*\lambda N}{8}}}.
\ee
Here we need to have $z'<0$, since the equation is obtained by squaring an equation whose left hand side is linear in $z'$ and has a positive coefficient, and whose right hand side is negative, after which we take the square root of $z'^2$. Since $R_0^2=2\lambda N \mu kR_*$, after defining
\be
  x=\mu^2r^2;\;\;\;
  y=\mu z;\;\;\;\;
  a=\mu R_0,
\ee
we obtain the equation
\be
  \frac{dy}{dx}=\pm\frac{\left(\frac{a^2}{x^2}-\frac{1}{2}+\frac{x}{4^2}\right)\sqrt{1-\frac{x}{4}}}{\sqrt{1-    
  \frac{x}{4}-\frac{x^3}{4^4}+\frac{x^2}{4^2}
  -\frac{a^2}{4^2}}}\,.
\ee
However, as explained in \cite{Lambert:2011eg}, we are in the approximation $a=\mu R_0\ll 1$, and the 
fixed sphere ground state solution is $r=R_0$, or
$x=a^2\ll 1$. That means that we can assume $x$ small in the above equation, thus
\be
  \frac{dy}{dx}\simeq -\frac{a^2}{2x^2}+\frac{1}{4}
\ee
which gives finally
\be
  z\simeq \frac{R_0^2}{2\mu r^2}+\frac{\mu r^2}{4}. \label{zsol}
\ee
This $z(r)$ has a minimum at 
\be
  r_*=\left(\frac{4\lambda N kR_*}{\mu}\right)^{1/4}=\left(\frac{\sqrt{2}R_0}{\mu}\right)^{1/2}\gg R_0,
\ee
and at $r_*$, the minimum value of $z$ is 
\be
  z_{min}=\sqrt{\mu\lambda NkR_*}=\frac{R_0}{\sqrt{2}}\,.
\ee
Note that here $R_0$ is written in physical spacetime variables, $R_{ph}^2=8\pi^2Nl_P^3f^2$, where $f=\sqrt{\mu k/(2\pi)}$ is the radius in the M2-brane worldvolume theory. 

To compare with the fuzzy funnel solutions (\ref{f2}), we note that $z$ is the equivalent of the M2-brane worldvolume direction $x_1$ and 
$r$ is the equivalent of the transverse direction $f(x_1)$. Therefore we must consider $r(z)$ but, since it has two branches, we must choose only one. The two branches are: $r(z)$ going from 0 to $r_*$ (for $z$ going from $\infty$ to $z_{min}$), and $r(z)$ going from $r_*$ to infinity. That 
would naively match the two solutions in (\ref{f2}), except for the fact that $r_*\gg R_0$, and $r_*\rightarrow \infty$ at $\mu\rightarrow 0$ (with 
$N$ very large), whereas $R_0=2\mu \lambda N\rightarrow 0$ as $\mu\rightarrow 0$. In fact, for $\mu\rightarrow 0$, we obtain from (\ref{zsol})
\be
  r(z)=\frac{R_0}{\sqrt{2\mu z}}\,.
\ee
This can be compared with the fuzzy funnel solution (\ref{fzfnl}), written as 
\be
  f(x_1)=\frac{f}{\sqrt{2\mu x_1}}\,,\label{fzfnl2}
\ee
so the spacetime solution is a {\it deformation} of the $\mu=0$ case. 
It also matches with the first solution in (\ref{f2}) in the $\mu\rightarrow 0$ limit. However if $\mu$ is fixed, the solution becomes  
(\ref{fzfnl2}) in the $x_1\rightarrow 0$ limit, corresponding to $z\rightarrow 0$, which is not even reachable by (\ref{zsol}). 

It is therefore unclear to us how to relate the two branches of (\ref{zsol}) to the two solutions of (\ref{f2}) precisely, other than through the general qualitative behaviour. We will leave a precise understanding of the matching to future work.

\section{Conclusions}
In this paper, we have studied various ans\"{a}tze for abelian reductions of the ABJM model, in the 
general case of nonzero mass, 
and used them to build a better defined AdS/CMT model. We have found a general 
abelianization ansatz (\ref{abelianmaster}, \ref{abelianpot}), 
using the matrices $G^\a$ that describe the fuzzy funnel BPS state and fuzzy sphere 
ground state, and that represent a consistent truncation. 
A further consistent truncation led to a model with topological vortex BPS solutions, 
but with $|\phi|\rightarrow 0$ at both $r=0$ and $r=\infty$ while yet another further consistent truncation 
led to a relativistic Landau-Ginzburg model which, depending on the parameter $c^2=\mu k/(2\pi)$ and 
on the scalar vev $b$, extrapolates between between the abelian-Higgs model, and a scalar $\phi^4$ theory. 

The second abelianization was used to take steps towards a better defined AdS/CMT 
model, since the ABJM model has a gravity dual, and the 
abelianization corresponds to the collective dynamics of ${\cal O}(N)$ out of the 
${\cal O}(N^2)$ fields. We also sketched a simple condensed matter 
model for a solid with free electrons that exhibits the same general features as 
the abelianization and leads to a bosonic Hubbard model, which 
in the continuum limit gives the relativistic Landau-Ginzburg system. It will 
be interesting to see if we can make more the model more concrete and elaborate further on 
its relation to ABJM. If successful, our construction provides, in our opinion, a concrete embedding
of the AdS/CMT correspondence in string theory.

In the last two sections, we studied various BPS solutions suggested by the abelianization, finding 
some generalizations of known solutions. We tried to find a spacetime 
interpretation for the BPS solutions in (\ref{f2}) as M2-M5 systems, 
with partial success. For small fluctuations we succeeded in matching this with the 
BIGGon solution for an M2 ending on a spherical M5, but for the full system 
we could only match only general qualitative behaviour and not 
the particular solution. It goes without saying that more work is needed to understand these solutions.

{\bf Acknowledgements}\\
We would like to thank Igor Barashenkov, Chris Clarkson, Aki Hashimoto, Andrey Pototskyy and Jonathan Shock for useful discussions at various stages of this work. HN would like to thank the University of Cape Town for hospitality 
during the time this project was started.
The work of HN is supported in part by CNPq grant 301219/2010-9. JM acknowledges support from the National Research Foundation (NRF) of South Africa under the Incentive Funding for Rated Researchers and Thuthuka programs. AM was supported by an NRF PhD scholarship and the University of Kartoum.

\end{document}